\documentclass[aps,prl,twocolumn,longbibliography,superscriptaddress]{revtex4-1}

\usepackage{graphicx}% Include figure files
\usepackage{epstopdf}
\usepackage{amsmath}
\usepackage{dcolumn}% Align table columns on decimal point
\usepackage{bm}% bold math
\usepackage{accents}
\usepackage{amssymb}
\usepackage{mathtools}
\usepackage{feynmp}
\usepackage{url}
\usepackage{setspace}
\usepackage{color}
\usepackage{pifont}
\usepackage{hyperref}
\usepackage{cleveref}

\usepackage{physics}

\begin{document}

\title{Quantum enhancement of information-mediated energy transfer}

\author{Shotaro Oki}
\affiliation{Department of Applied Physics, The University of Tokyo, 7-3-1 Hongo, Bunkyo-ku, Tokyo 113-8656, Japan}

\author{Yuki Kadono}
\affiliation{Department of Applied Physics, The University of Tokyo, 7-3-1 Hongo, Bunkyo-ku, Tokyo 113-8656, Japan}
\author{Kaito Tojo}
\affiliation{Department of Applied Physics, The University of Tokyo, 7-3-1 Hongo, Bunkyo-ku, Tokyo 113-8656, Japan}
\author{Takahiro Sagawa}
\affiliation{Department of Applied Physics, The University of Tokyo, 7-3-1 Hongo, Bunkyo-ku, Tokyo 113-8656, Japan}
\affiliation{Quantum-Phase Electronics Center (QPEC), The University of Tokyo, 7-3-1 Hongo, Bunkyo-ku, Tokyo 113-8656, Japan}
\affiliation{Inamori Research Institute for Science (InaRIS), Kyoto-shi, Kyoto 600-8411, Japan}
\author{Ken Funo}
\affiliation{Department of Applied Physics, The University of Tokyo, 7-3-1 Hongo, Bunkyo-ku, Tokyo 113-8656, Japan}

\date{July 2026}

\begin{abstract}
Thermodynamics of information identifies information flow as a thermodynamic resource, but whether quantum coherence and collective coupling can enhance it at low entropy-production cost remains unresolved. We address this question for interacting open quantum systems by deriving a thermodynamic uncertainty relation that bounds information flow in terms of entropy production in nonequilibrium steady states. We construct a quantum engine in which $N$-fold degenerate ground and excited states are collectively coupled to heat baths. Collective jumps enhance heat currents and information flow linearly with $N$, while entropy production remains independent of $N$, realizing a high-power, low-dissipation autonomous quantum Maxwell's demon that leverages collectively enhanced information flow to pump heat against a temperature gradient. Beyond steady states, collective interactions amplify the unitary component of quantum information flow, yielding a quadratic enhancement of the free-energy charging power of quantum batteries. Our results reveal scalable advantages of quantum coherence and collective effects in quantum engines.
\end{abstract}

\maketitle
\section{Introduction}
Over the past few decades, thermodynamics of information has clarified the fundamental thermodynamic costs of information processing and the performance limits of information-powered operations~\cite{NaturephysITD,Funo2018,PhysRevLett.100.080403,sagawaueda2009,PhysRevLett.104.090602,Sagawa2012PTP,PhysRevLett.109.180602,PhysRevLett.111.180603,PhysRevX.4.031015,PhysRevX.7.021003,QIF,yada2022}. 
By linking information-theoretic and thermodynamic quantities, this framework has shown that information can serve as a thermodynamic resource, enabling operations that go beyond the limitations of the conventional second law. Examples include work extraction beyond the free energy difference and heat-current generation against a thermal gradient. 
These ideas have been developed in both classical and quantum regimes, including information-powered refrigerators~\cite{Koski2015Refrigerator} and feedback cooling~\cite{Horowitz_2014,kumasaki2025}.
In particular, Maxwell's demons have been investigated in theoretical proposals~\cite{autodemonoriginal,PhysRevX.4.031015,PhysRevLett.111.230402,QIF} and experimental implementations~\cite{Naturephysdemon,PhysRevLett.117.240502,Cottet2017PNAS,Naghiloo2018PRL,Masuyama2018NatCommun,yada2025PRX}.

More recently, finite-time thermodynamic trade-off relations~\cite{shiraishisaito,PhysRevLett.121.070601,Funo_2019,PhysRevXVu} such as thermodynamic uncertainty relations (TURs)~\cite{NaturephysTUR,PhysRevLett.114.158101, PhysRevLett.116.120601,Universaltradeoffbetween,PhysRevLett.128.140602,PRXQuantum.6.010343} have provided a universal constraint on the precision of currents in terms of the entropy production. TURs have been extended to information-processing settings in classical systems~\cite{PhysRevE.100.052137,Vu_2020,OtsuboItoDechantSagawa2020,TURclassicaldemon} and in open quantum systems~\cite{Vu2026,tojo2026thermodynamicuncertaintyrelationcontinuous,HonmaVu2026}.

Collective quantum enhancement provides a route to mitigate the constraints imposed by thermodynamic trade-off relations. These trade-off relations show that the squared heat current is bounded above by the product of the entropy production and the activity, which quantifies the average jump rate of the system. When multiple quantum systems are coherently superposed and collectively coupled to a bath, the resulting average jump rate can be enhanced, as exemplified by Dicke superradiance~\cite{Dicke1954}. This enhancement relaxes the trade-off relation: by sufficiently increasing the activity, the bound on the heat current is increased, enabling larger heat currents even when the entropy production remains fixed~\cite{TajimaFuno,FunoTajima}. Similar collective effects have also been explored in heat engines~\cite{SuperradiantQuantumHeatEngine,Niedenzu_2018,quantumenhancedheatengine,yoshimura2026}, photocells~\cite{Photocell}, Landauer's principle~\cite{landauer}, and quantum batteries~\cite{RevModPhys.96.031001,enhancingquantumbattery,highpowercollective,PhysRevLett.134.130401}.

These developments raise a natural question: can collective quantum enhancement also enhance information flow? Despite the progress described above, it remains unclear whether information flow can be enhanced without a comparable increase in entropy production. Establishing such a mechanism would open a route to high-power, low-dissipation processes of information-mediated energy transfer, in which collectively enhanced information flow serves as a scalable thermodynamic resource.

\begin{figure}
    \centering
    \includegraphics[width=\linewidth]{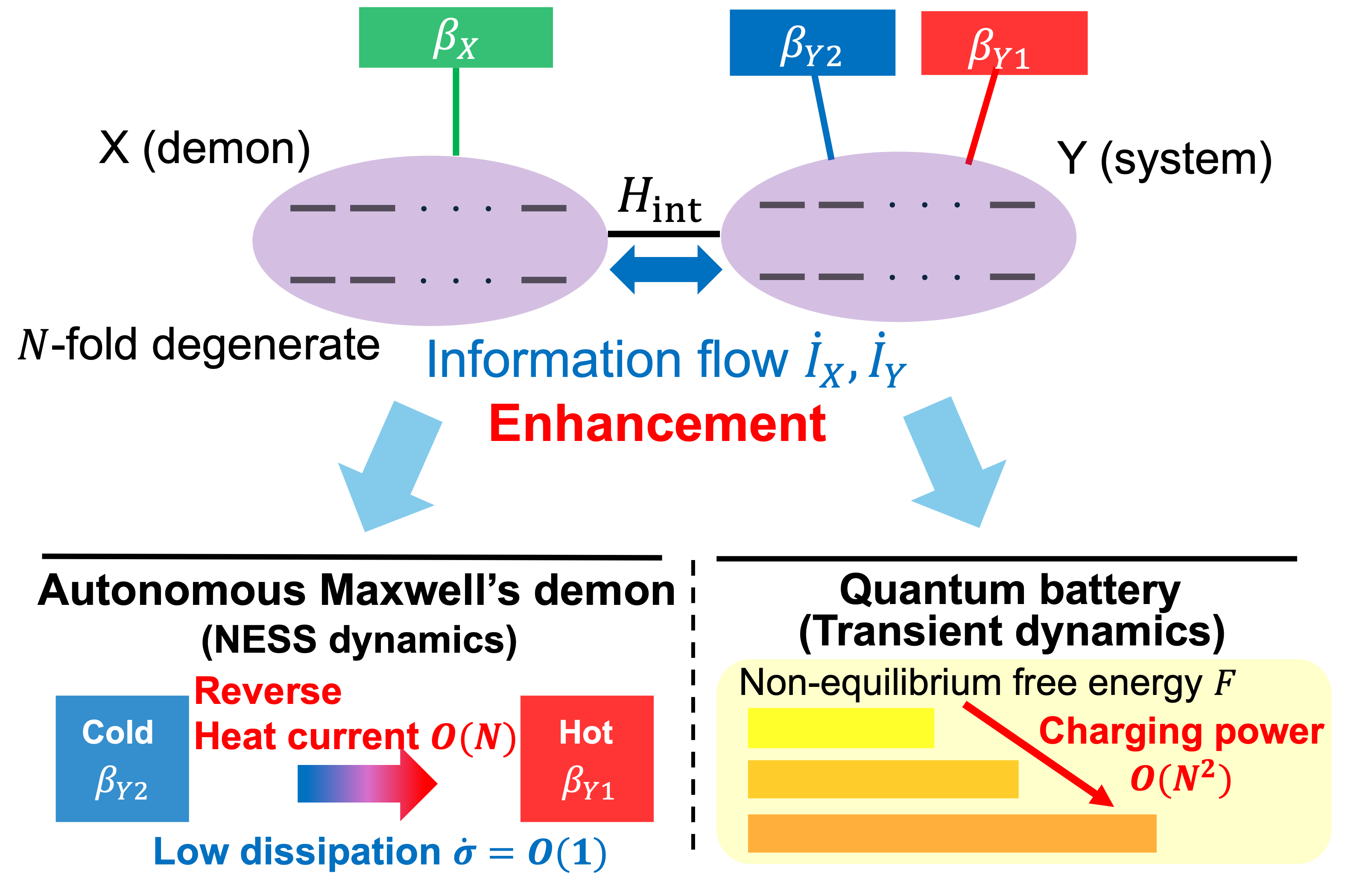}
    \caption{Conceptual overview of the information-flow enhancement. Two $N$-fold degenerate subsystems, $X$ and $Y$, are coupled by $H_{\rm int}$ and exchange quantum information flow. Quantum coherence and collective dynamics enhance this information flow, leading to two consequences: an autonomous Maxwell's demon that pumps an $O(N)$ reverse heat current with $O(1)$ entropy production in the NESS, and a protocol for charging a quantum battery with free-energy charging power scaling as $O(N^2)$.
    }
    \label{fig:main}
\end{figure}

Here, we address this question in a setup consisting of two interacting open quantum systems, as schematically shown in Fig.~\ref{fig:main}. We derive a short-time TUR that explicitly relates information flow to the partial entropy production in the nonequilibrium steady state (NESS). We then use this trade-off to identify the conditions under which collective enhancement is possible and construct a concrete setup that realizes the resulting scaling.
By introducing a system with $N$-fold degenerate ground and excited states collectively coupled to heat baths, we show that the heat currents and information flow are both enhanced to $O(N)$, while the partial entropy production rates remain at $O(1)$. The resulting $O(N)$ information flow acts as a thermodynamic resource that drives an $O(N)$ apparent violation of the conventional second law for the subsystem. This realizes an autonomous quantum Maxwell's demon that produces an $O(N)$ reverse heat current against a thermal gradient while the total entropy production remains $O(1)$.
Finally, we extend our analysis beyond the NESS to transient regimes, where the unitary contribution to the information flow becomes relevant. We show that collective interactions between the subsystems can amplify the unitary component of information flow to $O(N^2)$. Applying this mechanism to nonequilibrium free-energy charging of a quantum battery, we demonstrate an $O(N^2)$ enhancement of the free-energy charging power.

\section{Results}
\subsection{Setup}
We consider an open quantum system composed of two interacting subsystems, denoted by $X$ and $Y$. The Hamiltonian of the composite system is given by
\begin{equation}
    H_{\rm sys}=H_X +H_Y+H_{\rm{int}},
\end{equation}
where $H_X$ and $H_Y$ are the Hamiltonians of the two subsystems, $H_{\rm int}$ is their interaction, and all terms are time independent. Each subsystem is weakly coupled to its own set of independent heat baths: subsystem $X$ to baths $\{B_{X_i}\}$ and subsystem $Y$ to baths $\{B_{Y_i}\}$ at inverse temperatures $\beta_{X_i}$ and $\beta_{Y_i}$, respectively (see Fig.~\ref{fig:main}).

Under the standard Born--Markov--secular approximation, the time evolution of the system's density matrix $\rho$ is described by a Gorini--Kossakowski--Sudarshan--Lindblad (GKSL) master equation~\cite{BreuerPetruccione} (setting $\hbar = 1$):
\begin{align}\label{GKSL}
    \dot{\rho} =\mathcal{L}[\rho]\coloneqq&-i[H_{\rm sys}, \rho] + \mathcal{D}(\rho).
\end{align}
Here, 
$\mathcal{D}(\rho) = \mathcal{D}_X(\rho) + \mathcal{D}_Y(\rho)$ is the dissipator, and
\begin{align}
    \mathcal{D}_{X}(\rho)&\coloneqq\sum_{i}\mathcal{D}_{X_i}(\rho)\notag\\
    &= \sum_{i,\omega} \gamma_{X_i}(\omega) \left( L_{X_i}^{\omega} 
    \rho \left(L_{X_i}^{\omega}\right)^\dagger- \frac{1}{2} 
    \left\{\left(L_{X_i}^{\omega}\right)^\dagger L_{X_i}^{\omega}, 
    \rho\right\} \right)
\end{align}
describes the effect of the baths $\{B_{X_{i}}\}$ coupled to subsystem $X$, and $\mathcal{D}_{Y}(\rho)$ is defined in an analogous manner. Here, $L_{X_i}^{\omega}$ is the quantum jump operator induced by the $X_i$-th heat bath, which lowers the system energy by $\omega$, i.e., $[H_{\rm sys},L_{X_i}^{\omega}]=-\omega L_{X_i}^{\omega}$. The coefficient $\gamma_{X_i}(\omega) \geq 0$ is the corresponding transition rate, satisfying the local detailed balance condition $\frac{\gamma_{X_i}(-\omega)}{\gamma_{X_i}(\omega)}=e^{-\beta_{X_i}\omega}$.

We now introduce a thermodynamic framework for this open quantum system~\cite{QIF}. The total entropy production rate is defined as $\dot{\sigma} \coloneqq \dot{S}_{\rm{tot}} - \sum_{i}\beta_{X_i}\dot{Q}_{X_i} - \sum_{i}\beta_{Y_i}\dot{Q}_{Y_i}$, where $\dot{Q}_{X_i} \coloneqq \Tr[\mathcal{D}_{X_i}(\rho)\,H_{\rm sys}]$ is the heat current from the $X_{i}$-th bath into the system, and $\dot{Q}_{Y_i}$ is defined analogously for subsystem $Y$. This total entropy production rate is decomposed as $\dot{\sigma}=\dot{\sigma}_X+\dot{\sigma}_Y$, where the partial entropy production rate for subsystem $X$ is defined as~\cite{QIF}
\begin{equation}\label{partialentropy}
    \dot{\sigma}_X \coloneqq \dot{S}_X - \sum_{i}\beta_{X_i}\dot{Q}_{X_i} 
    - \dot{I}_X \geq 0,
\end{equation}
where $\dot{S}_X \coloneqq -\Tr[\dot{\rho}_X \ln \rho_X]$ denotes the rate of change of the von Neumann entropy of the subsystem $X$. The partial entropy production rate $\dot{\sigma}_Y$ is defined analogously. The information flow $\dot{I}_X$ is defined as~\cite{QIF}
\begin{align}\label{QIFdef}
    \dot{I}_X &\coloneqq -\Tr[\dot{\rho}_X \ln \rho_X] 
    + \Tr[\mathcal{D}_{X}[\rho]\ln\rho].
\end{align}
This quantity represents the contribution of subsystem $X$ to the rate of change of the quantum mutual information ${I}_{XY} = S_X + S_Y - S_{\rm{tot}}$, as $\dot{{I}}_{XY} = \dot{I}_X + \dot{I}_Y$~\cite{QIF}. A positive information flow identifies the subsystem that acquires information, whereas a negative information flow identifies the subsystem subjected to feedback control. In the setup of the autonomous Maxwell's demon considered below, subsystem $X$ satisfies $\dot I_X\geq0$ and acts as the demon, whereas subsystem $Y$ satisfies $\dot I_Y\leq0$ and acts as the feedback-controlled system.

The non-negativity of the partial entropy production rate $\dot{\sigma}_X \geq 0$ (and similarly $\dot{\sigma}_Y \geq 0$) represents the second law of thermodynamics for subsystems with quantum information flow. Rearranging the inequality $\dot {\sigma}_Y\geq0$ gives
\begin{equation}\label{secondlawY}
    \dot{S}_Y - \sum_i\beta_{Y_i}\dot{Q}_{Y_i}\geq \dot{I}_Y.
\end{equation}
In the case without information processing, where the information flow is absent, the left-hand side of Eq.~\eqref{secondlawY} corresponds to the usual entropy production rate, i.e., the entropy change of subsystem $Y$ and the baths $\{B_{Y_{i}}\}$, and is therefore non-negative. In contrast, when $\dot{I}_Y\leq0$, the left-hand side of Eq.~\eqref{secondlawY} can become negative, leading to an apparent violation of the second law. The magnitude of this violation is set by $|\dot I_Y|$, which serves as a thermodynamic resource for information-driven operations such as Maxwell's demon. 

\subsection{Short-time TUR for quantum information flow}
\begin{figure*}
    \centering
    \includegraphics[width=\linewidth]{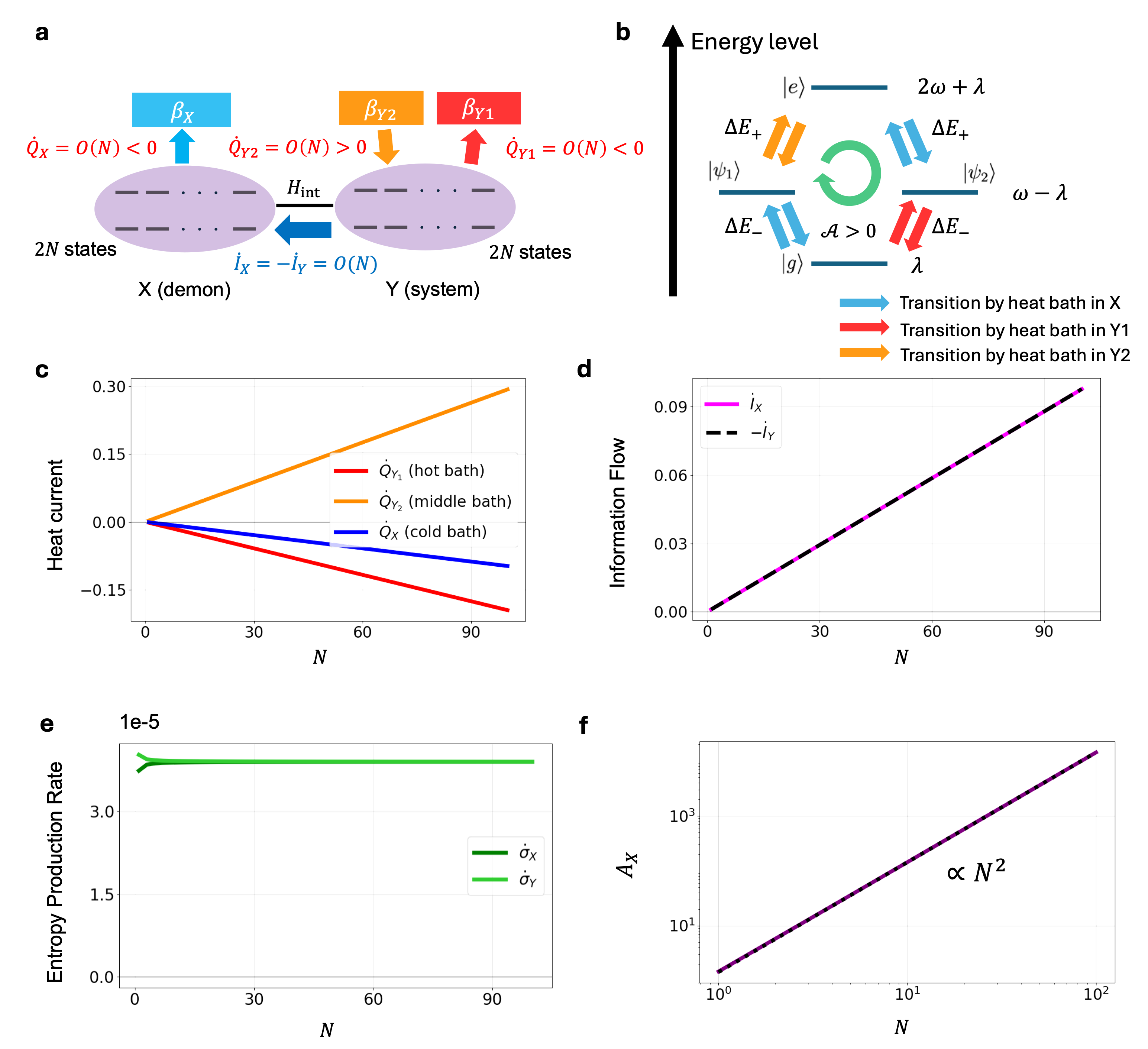}
    \caption{Setup and numerical verification of the $O(N)$ scaling in the autonomous Maxwell's demon.
    \textbf{(a)} Schematic of the bipartite system. Subsystem $X$ acts as the demon and is coupled to a cold bath at inverse temperature $\beta_X$, while subsystem $Y$ is coupled to two baths at $\beta_{Y_1}$ and $\beta_{Y_2}$. The demon generates an $O(N)$ information flow, $\dot{I}_X=-\dot{I}_Y=O(N)$, which drives an $O(N)$ reverse heat current in subsystem $Y$ against the thermal gradient, while $\dot{\sigma}_{\rm tot}=\dot{\sigma}_{X}+\dot{\sigma}_{Y}$ remains at $O(1)$.
    \textbf{(b)} Energy-level diagram of the four collective eigenstates $\ket{g}$, $\ket{\psi_1}$, $\ket{\psi_2}$, and $\ket{e}$, together with the thermodynamic cycle driven by baths $X$, $Y_1$, and $Y_2$.
    \textbf{(c)} Heat currents $\dot{Q}_{Y_1}$, $\dot{Q}_{Y_2}$, and $\dot{Q}_X$ as functions of $N$, confirming the $O(N)$ reverse heat current from the colder bath $Y_2$ to the hotter bath $Y_1$.
    \textbf{(d)} Information flows $\dot{I}_X$ and $\dot{I}_Y$, both scaling linearly with $N$.
    \textbf{(e)} Partial entropy production rates $\dot{\sigma}_X$ and $\dot{\sigma}_Y$ scaling as $O(1)$.
    \textbf{(f)} Partial activity $A_{X}$, scaling quadratically with $N$. 
    }
    \label{fig:demoncalc}
\end{figure*}
We first derive a short-time TUR that explicitly incorporates quantum information flow to investigate whether information flow can be enhanced while the partial entropy production remains small. For subsystem $X$, we establish that the square of the total heat current is bounded by the partial entropy production rate $\dot{\sigma}_X$ and the partial activity $A_X$~\cite{shiraishisaito,TajimaFuno,FunoTajima}:
\begin{equation}\label{TUR}
    \left(\sum_{i}\dot{Q}_{X_i}\right)^2 \leq \frac{1}{2}\dot{\sigma}_X A_X,
\end{equation}
with an analogous expression for subsystem $Y$. Here, $A_X=\sum_{i,\omega }\omega^2\gamma_{X_i}(\omega)\Tr\!\left[(L_{X_i}^\omega)^\dagger L_{X_i}^\omega \rho\right] $ quantifies the average jump rate. We note that this inequality is consistent with the short-time limit of the TUR obtained in~\cite{Vu2026}, as the fluctuation of heat current reduces to $A_X$ in this limit.

We next apply Eq.~\eqref{TUR} to derive a bound on the information flow in the NESS, i.e., $\mathcal{L}[{\rho}_{\rm ss}] = 0$.  We assume that subsystem $X$ is coupled to a single heat bath at inverse temperature $\beta_X$ and serves as a demon ($\dot I_X\geq 0$). Using $\dot S_X=0$ in the NESS, Eq.~\eqref{partialentropy} implies $-\beta_X\dot Q_X\geq \dot I_X$. Combining this relation with Eq.~\eqref{TUR} yields
\begin{equation}\label{ITUR}
    |\dot{I}_X|^2 \leq \frac{1}{2}\beta_X^2\,\dot{\sigma}_X A_X.
\end{equation}
We refer to this as the \textit{short-time information TUR}, which constrains the magnitude of the information flow in terms of the partial entropy production and $A_{X}$. In what follows, we discuss to what extent the information flow can be enhanced by exploiting quantum effects based on the bound~\eqref{ITUR}.

\subsection{Enhancement of quantum information flow}
Let $N$ denote the degeneracy of each subsystem. Previous studies showed that collective quantum jumps can enhance the partial activity from the conventional additive scaling $A_X=O(N)$ to the quadratic scaling $A_X=O(N^2)$~\cite{TajimaFuno,FunoTajima}. 
Equation~\eqref{ITUR} then shows that a linear enhancement of information flow is achievable while keeping the entropy production rate at $O(1)$: if $A_{X}=O(N^{2})$ and $\dot{\sigma}_{X}=O(1)$, the upper bound on $|\dot{I}_{X}|$ scales as $O(N)$. 
A similar argument applies to subsystem $Y$. Since $\dot{I}_{Y}=-\dot{I}_{X}$ in the NESS, we have $|\dot{I}_{Y}|=O(N)$. Applying Eq.~\eqref{TUR} to subsystem $Y$, this scaling is compatible with $\dot{\sigma}_{Y}=O(1)$ when the corresponding partial activity also scales as $A_{Y}=O(N^{2})$ (see Methods). Thus, the collective setup allows
\begin{align}
    \dot{\sigma}_{X}=O(1), \quad \dot{\sigma}_{Y}=O(1),\quad  \dot{I}_{X}=-\dot{I}_{Y}=O(N),
\end{align}
and enables an $O(N)$ apparent violation of the conventional second law~\eqref{secondlawY}.

The preceding argument shows only that these scalings are compatible with the TURs. However, in what follows, we explicitly construct a model of an autonomous Maxwell's demon that satisfies the above scalings. Specifically, we show that an $O(N)$ information flow mediates an $O(N)$ heat current against a thermal gradient with $O(1)$ dissipation. Demonstrating this scaling in an information-powered heat engine is a central result of this paper.

\subsection{Autonomous Maxwell's demon}
To realize an autonomous Maxwell's demon~\cite{autodemonoriginal,PhysRevX.4.031015,QIF} and identify the mechanism responsible for the $O(N)$ scaling, we consider two subsystems, $X$ and $Y$, each with $N$-fold degenerate ground and excited states~\cite{TajimaFuno} (see Fig.~\ref{fig:main}). The Hamiltonians of the subsystems are given by
\begin{equation}
    H_X = \sum_{i_X=1}^N \omega \ket{e, i_X}\bra{e, i_X}, \quad 
    H_Y = \sum_{j_Y=1}^N \omega \ket{e, j_Y}\bra{e, j_Y},
\end{equation}
where $\omega$ is the excitation energy. Here, $\ket{e, i_X}$ denotes the $i_X$-th $N$-fold degenerate excited state of subsystem $X$, and the corresponding ground state is denoted by $\ket{g, i_X}$. We define the collective superposition of the ground and excited states of subsystem $X$ as $\ket{g,X}\coloneqq\frac{1}{\sqrt{N}}\sum_{i_X}\ket{g,i_X}$, $\ket{e,X}\coloneqq\frac{1}{\sqrt{N}}\sum_{i_X}\ket{e,i_X}$ and analogously for subsystem $Y$.
The interaction Hamiltonian is given by
\begin{align}\label{bipint}
    H_{\rm{int}}
    &= \lambda \sum_{i=1}^{N} \sum_{j=1}^{N}
    \sigma_z^{X_{i}} \sigma_z^{Y_{j}},
\end{align}
where $\sigma_z^{X_{i}} = -\ket{g,i_X}\bra{g,i_X} 
+ \ket{e,i_X}\bra{e,i_X}$, and $\sigma_z^{Y_{j}}$ denotes the analogous operator for subsystem $Y$. The interaction Hamiltonian commutes with the Hamiltonians of the subsystems:
\begin{equation}\label{bipcond}
    [H_{\rm int}, H_X] = 0, \qquad [H_{\rm int}, H_Y] = 0.
\end{equation}
We refer to this property as the \textit{bipartite condition}, following the framework of autonomous Maxwell's demon~\cite{QIF}. When Eq.~\eqref{bipcond} is satisfied, the classical limit of this model reproduces a standard classical autonomous Maxwell's demon model~\cite{QIF,PhysRevX.4.031015,autodemonoriginal}. Therefore, our model naturally extends this concept to the quantum regime.

Figure~\ref{fig:demoncalc} illustrates the setup and the resulting energy-level structure.
The four collective eigenstates of $H_{\rm sys}$ are $ \ket{g} \coloneqq \ket{g,X}\ket{g,Y}, \ket{e} \coloneqq \ket{e,X}\ket{e,Y}, \ket{\psi_1} \coloneqq \ket{e,X}\ket{g,Y}, \ket{\psi_2} \coloneqq \ket{g,X}\ket{e,Y}$, with eigenvalues  $\lambda$, $2\omega+\lambda$, $\omega-\lambda$, and $\omega-\lambda$, respectively. 
For collective system--bath couplings, the relevant downward jump operators take the form
\begin{align}\label{explicitjump}
    L_{X}^{\Delta E_-} &= N\ket{g}\bra{\psi_1}, 
    &
    L_{X}^{\Delta E_+} &= N\ket{\psi_2}\bra{e}, \nonumber\\
    L_{Y_1}^{\Delta E_-} &= N\ket{g}\bra{\psi_2}, 
    &
    L_{Y_2}^{\Delta E_+} &= N\ket{\psi_1}\bra{e},
\end{align}
where $\Delta E_\pm \coloneqq \omega\pm 2\lambda>0$. The derivation of these collective jump operators is given in Methods. Their amplitudes scale as $N$, yielding $A_X,A_Y=O(N^2)$~\cite{TajimaFuno,FunoTajima}.

For this system to operate as an autonomous Maxwell's demon, subsystem $X$ is coupled to a single cold bath at inverse temperature $\beta_X$, while subsystem $Y$ is coupled to two baths $Y_1$ and $Y_2$ with $\beta_X > \beta_{Y_2} > \beta_{Y_1}$. We assume energy-selective couplings such that bath $Y_1$ couples to the smaller energy gap transition $\Delta E_-$, and bath $Y_2$ couples to the larger energy gap transition $\Delta E_+$. In the NESS, transitions among these four states form a closed thermodynamic cycle, as depicted in Fig.~\ref{fig:demoncalc}b. The population dynamics of this model are described by a Markov jump process, to which Schnakenberg's network theory~\cite{network} applies. We define the positive cycle affinity as $\mathcal A=\beta_{X}(\Delta E_{+}-\Delta E_{-})+\beta_{Y_{1}}\Delta E_{-}-\beta_{Y_{2}}\Delta E_{+}>0$, which fixes the cycle direction corresponding to the operation of Maxwell's demon, and we set $\mathcal{A}=O(1/N)$.
Under this condition, the collective transition rates drive the full four-state cycle and generate an $O(N)$ information flow, $\dot I_X=-\dot I_Y=O(N)$. This information flow acts as a thermodynamic resource that sustains an $O(N)$ reverse heat current from the colder bath $Y_2$ to the hotter bath $Y_1$, as numerically confirmed in Fig.~\ref{fig:demoncalc}c and d. At the same time, the partial entropy production rates remain $\dot\sigma_X=O(1)$ and $\dot\sigma_Y=O(1)$, as shown in Fig.~\ref{fig:demoncalc}e. Thus, the model realizes a high-power, low-dissipation autonomous information engine. This scaling goes beyond the conventional (non-collective) case, for which $A_{X}=O(N)$ and the partial entropy production should scale at least as $O(N)$ to realize an $O(N)$ information flow. These results demonstrate a quantum scaling advantage for information-thermodynamic engines.

\subsection{Analysis of information flow away from the steady state} 
We now go beyond the NESS and analyze the transient dynamics of the information flow defined in Eq.~\eqref{QIFdef}. Following Ref.~\cite{QIF}, we decompose the information flow into two contributions (see Methods):
\begin{align}\label{decomposition}
    \dot{I}_Y &= \underbrace{\Tr[i[H_{\rm{int}},\rho]
    \ln\rho_Y]}_{\dot I_Y^{\rm uni}}
    + \underbrace{\Tr[\mathcal{D}_Y[\rho]\ln\rho]
    -\Tr[\mathcal{D}[\rho]\ln\rho_Y]}_{\dot{I}_Y^{\rm{diss}}}.
\end{align}
The first term, $\dot{I}_Y^{\rm uni}$, originates from the unitary evolution generated by $H_{\rm int}$ and vanishes in the NESS considered here. In contrast, the second and third terms together constitute the dissipative contribution $\dot{I}_Y^{\rm{diss}}$, arising from the coupling to the heat baths.

\subsection{Scaling of unitary information flow}
In the analysis of the NESS above, the information flow is purely dissipative, and its enhancement originates from collective jumps induced by the heat baths. Beyond the NESS, the unitary contribution $\dot I_Y^{\rm uni}$ is generated by the interaction Hamiltonian $H_{\rm int}$. Thus, collective interactions in $H_{\rm int}$ can enhance the unitary information flow (see also Methods for a quantitative argument based on the spectral width of $H_{\rm int}$). In the autonomous Maxwell's demon setup, the interaction Hamiltonian in Eq.~\eqref{bipint} does not contain collective interactions, yielding only an $O(1)$ unitary information flow. By contrast, we consider the following collective interaction 
\begin{align}\label{bipint2}
    H_{\rm{int}}^{B}
    &= \lambda \sum_{i,i'=1}^{N} \sum_{j,j'=1}^{N}
    \sigma_z^{X_{ii'}} \sigma_z^{Y_{jj'}}\nonumber\\
    &=\lambda N^2(-\ket{g,X}\bra{g,X}+\ket{e,X}\bra{e,X})\nonumber\\&\quad\otimes(-\ket{g,Y}\bra{g,Y}+\ket{e,Y}\bra{e,Y}),
\end{align}
where $\sigma_z^{X_{ii'}} = -\ket{g,i_X}\bra{g,i'_X} + 
\ket{e,i_X}\bra{e,i'_X}$ and $\sigma_z^{Y_{jj'}}$ is defined analogously. This collective interaction enhances the effective interaction strength to $\lambda N^2$, where we consider the range of $N$ satisfying $\omega>2\lambda N^2$, thereby allowing the unitary information flow to scale as $O(N^2)$.

An $O(N^2)$ enhancement can also be realized beyond the bipartite condition. As an example, we consider a Jaynes--Cummings-type interaction,
\begin{align}\label{nonbipint}
    H_{\rm{int}}^{NB}
    &= g \sum_{i,i'=1}^N \sum_{j,j'=1}^N
    \left(
    \sigma_+^{X_{ii'}} \sigma_-^{Y_{jj'}}
    + \sigma_-^{X_{ii'}} \sigma_+^{Y_{jj'}}
    \right)\nonumber\\
    &=gN^2(\ket{e,X}\ket{g,Y}\bra{g,X}\bra{e,Y}\nonumber\\&\quad+\ket{g,X}\ket{e,Y}\bra{e,X}\bra{g,Y}).
\end{align}
Here, $\sigma_+^{X_{ii'}}=\ket{e,i'_X}\bra{g,i_X},$ 
$\sigma_-^{X_{ii'}}=\ket{g,i_X}\bra{e,i'_X}$, with analogous definitions for subsystem $Y$. We restrict the parameter range to $\omega>gN^2$. This interaction does not satisfy the bipartite condition given in Eq.~\eqref{bipcond}, yet similarly achieves an effective interaction strength that scales as $O(N^2)$, allowing an $O(N^2)$ enhancement of the unitary information flow.

\subsection{Transient dynamics and quantum battery charging}
We now apply the $O(N^2)$ enhancement of the unitary information flow to a quantum battery charging protocol. In this setting, subsystem $X$ acts as the charger and subsystem $Y$ as the battery. Each subsystem is coupled to a single heat bath at the same inverse temperature, $\beta_X=\beta_Y = \beta$. We consider the weak-dissipation regime in which $\gamma_{X}(\omega)$ and $\gamma_{Y}(\omega)$ are small compared with the strength of the interaction Hamiltonian. In this regime, the unitary component of the information flow dominates during the battery charging protocol.
To quantify the thermodynamic resource stored in subsystem $Y$, we define its excess free energy as 
\begin{align}
\mathcal{W}_{Y}(t):=F_{Y}(t)-F_{Y}^{\rm eq},
\end{align}
where $F_{Y}(t)\coloneqq \Tr[H_Y\rho_Y(t)]-\beta^{-1}S(\rho_Y(t)) $ is the nonequilibrium free energy at inverse temperature $\beta$, and $F_{Y}^{\rm eq}=-\beta^{-1}\ln Z_Y, Z_{Y}\coloneqq\Tr[e^{-\beta H_Y}]$ is the equilibrium free energy. In the standard setting where $Y$ is coupled to a single heat bath at inverse temperature $\beta$, the maximum average work extractable from $\rho_{Y}(t)$ is given by $\mathcal{W}_{Y}(t)$~\cite{batteryNFE,shi2022,Kamin_2023}. The net gain in stored resource generated during the charging process is then $\Delta \mathcal{W}_{Y}(t):=\mathcal{W}_{Y}(t)-\mathcal{W}_{Y}(0)=F_{Y}(t)-F_{Y}(0)$. We define the average free-energy charging power as  $P_{Y}=\Delta \mathcal{W}_{Y}(t^{*})/t^{*}$, where $t^{*}$ is the earliest time at which $\Delta \mathcal{W}_{Y}(t)$ reaches its first maximum.

Using the decomposition of the information flow~\eqref{decomposition}, the time derivative of $F_Y$ can be written as
\begin{align}
    \dot{F}_Y
    &=
    \dot{J}_Y^{\rm uni}
    -
    \beta^{-1}\dot{I}_Y^{\rm uni}
    \nonumber\\&\quad+
    \Tr[\mathcal{D}[\rho]H_Y]
    -
    \beta^{-1}
    \left(
    \dot{\sigma}_Y+\dot{I}_Y^{\rm diss}
    \right)
    -
    \dot{Q}_Y,
\end{align}
where $\dot{J}_Y^{\rm uni}=\Tr\left[-i[H_{\rm int},\rho]H_Y\right]$ is the unitary energy flow into subsystem $Y$ generated by the interaction Hamiltonian. 
In the weak-dissipation regime, dissipative contributions to $\dot F_Y$ are subleading, so that the free energy change is dominated by the unitary energy and information flow, $\dot F_Y\simeq \dot J_Y^{\rm uni}-\beta^{-1}\dot I_Y^{\rm uni}$. Under the bipartite condition, we have $\dot{J}_Y^{\rm uni}=0$, so the free-energy change is driven solely by the unitary information flow. In contrast, in the non-bipartite case, both $\dot{I}_Y^{\rm uni}$ and $\dot{J}_Y^{\rm uni}$ contribute to the charging dynamics. We analyze these two cases separately below.

\subsection{Bipartite case}
\begin{figure*}
    \centering
    \includegraphics[width=\linewidth]{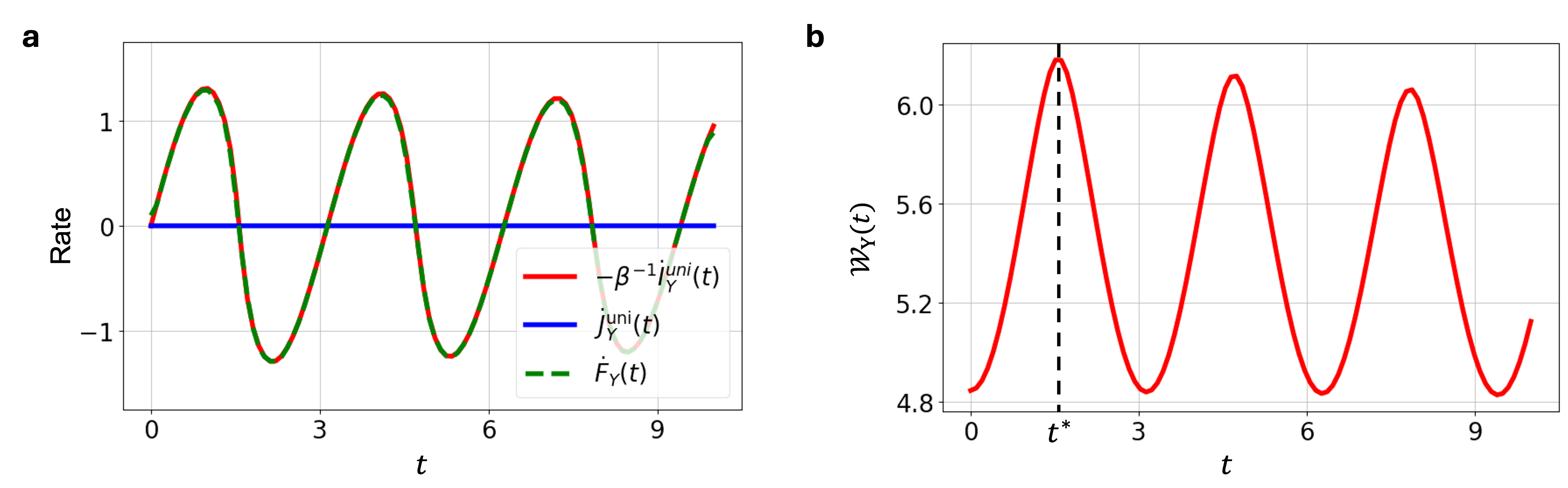}
    \caption{Transient dynamics of the bipartite model.
    \textbf{(a)}~Time evolution of the free energy rate $\dot{F}_Y$ and its contributions: the unitary information flow term $-\beta^{-1}\dot{I}^{\rm{uni}}_{Y}(t)$ (red) and $\dot{F}_Y(t)$ (green dashed) coincide, while the energy flow $\dot{J}_Y^{\rm{uni}}(t)$ (blue) vanishes identically under the bipartite condition. 
    \textbf{(b)}~Time evolution of $ \mathcal{W}_Y(t)$ (red), showing coherent oscillations driven purely by the unitary information flow. The vertical dashed line indicates the charging time $t^*$, defined as the time at which $\mathcal{W}_Y(t)$ is maximized.}
    \label{fig:bipartitetrans}
\end{figure*}

\begin{figure*}
    \centering
    \includegraphics[width=\linewidth]{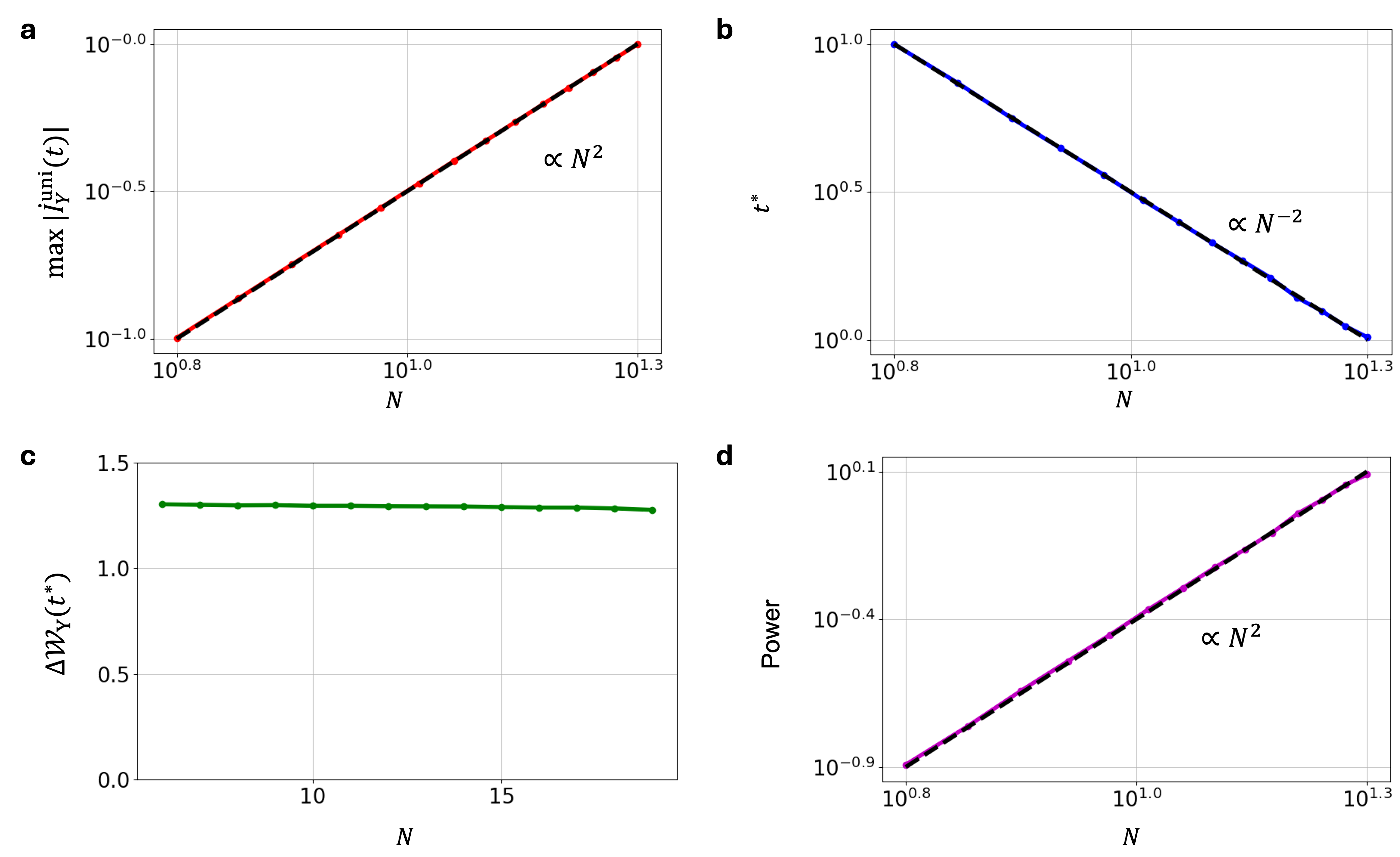}
    \caption{Scaling behavior of the bipartite quantum battery.
    \textbf{(a)}~Maximum unitary information flow $\max|\dot{I}^{\rm{uni}}_{Y}|$ as a function of $N$, showing $O(N^2)$ growth consistent with the collective 
    enhancement. 
    \textbf{(b)}~Charging time $t^*$ as a function of $N$ (see also Fig. 3b). The charging time decreases as $O(1/N^2)$.
    \textbf{(c)}~$\Delta\mathcal{W}_Y(t^*)$ (green) remains $O(1)$.
    \textbf{(d)}~Average charging power $P_Y=\Delta\mathcal{W}_Y(t^*)/t^*$, showing $O(N^2)$ growth.}
    \label{fig:biptransscale}
\end{figure*}

We first consider the bipartite model described by Eq.~\eqref{bipint2}. In this model, we have $\dot{J}_Y^{\rm uni}=0$, and $\dot{F}_{Y}\simeq -\beta^{-1}\dot{I}_{Y}^{\rm uni}$.
We initialize the system in the classically correlated state
\begin{align}\label{bipini}
    \rho(0)
    &=
    \frac{1}{2}
    \ket{g,X}\bra{g,X}
    \otimes
    \ket{+,Y}\bra{+,Y} \nonumber\\&\quad+
    \frac{1}{2}
    \ket{e,X}\bra{e,X}
    \otimes
    \ket{-,Y}\bra{-,Y},
\end{align}
where $\ket{\pm,Y}=\frac{1}{\sqrt{2}}(\ket{g,Y}\pm\ket{e,Y})$. The resulting dynamics are shown in Fig.~\ref{fig:bipartitetrans}. At early times, the nonequilibrium free energy $F_Y(t)$ increases. The numerical results show that $\dot{J}_Y^{\rm uni}=0$, and that the increase of $F_Y(t)$ is driven purely by the unitary information flow. The scaling behavior is summarized in Fig.~\ref{fig:biptransscale}. The maximum unitary information flow grows as $O(N^2)$, while $\Delta\mathcal{W}_Y(t^*)$ remains $O(1)$. The charging time $t^*$ decreases as $O(1/N^2)$, and the average charging power $P_Y$ increases as $O(N^2)$, demonstrating fast free-energy charging with a finite stored free-energy gain.

This process can be interpreted as an autonomous coherent feedback operation~\cite{coherentfeedback}. The initial classical correlation with subsystem $X$ determines the conditional unitary operation applied to subsystem $Y$, and this feedback is implemented autonomously by the interaction Hamiltonian. Through this process, the interaction converts initial correlations with the charger into coherence in the battery, thereby increasing the nonequilibrium free energy.

\subsection{Non-bipartite case}
We next consider the non-bipartite model~\eqref{nonbipint}, where both the unitary information flow and the unitary energy flow contribute to the charging dynamics. We initialize the system as
\begin{equation}\label{nonbipini}
    \rho(0)
    =
    \ket{e,X}\bra{e,X}
    \otimes
    \frac{1}{2}
    \left(
    \ket{g,Y}\bra{g,Y}
    +
    \ket{e,Y}\bra{e,Y}
    \right),
\end{equation}
where subsystem $X$ is initially excited and subsystem $Y$ is in a mixed state.

\begin{figure*}
    \centering
    \includegraphics[width=\linewidth]{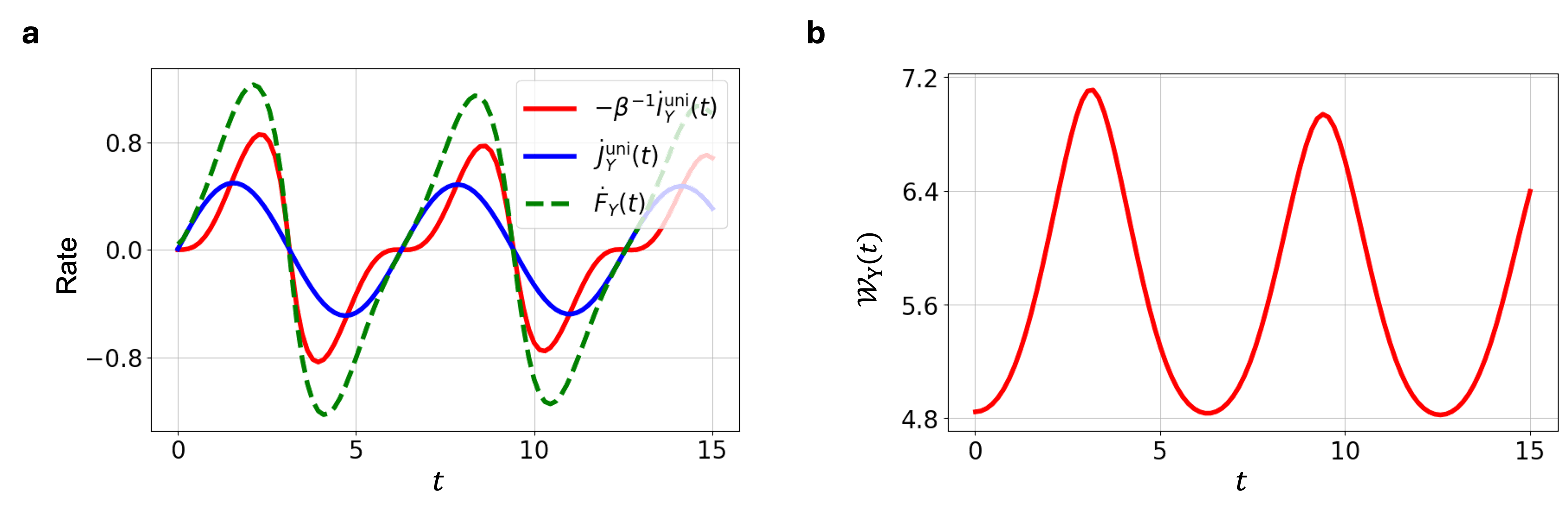}
    \caption{Transient dynamics of the non-bipartite model.
    \textbf{(a)}~Time evolution of the free energy rate $\dot{F}_Y$ and its contributions: unlike the bipartite case, both the unitary information flow $-\beta^{-1}\dot{I}_{Y}^{\rm uni}(t)$ (red) 
    and the energy flow $\dot{J}_Y^{\rm{uni}}(t)$ (blue) are nonzero and together drive $\dot{F}_Y(t)$ (green dashed).
    \textbf{(b)}~Time evolution of $ \mathcal{W}_Y(t)$ (red), showing coherent oscillations.}
    \label{fig:nonbipartitetrans}
\end{figure*}

As shown in Fig.~\ref{fig:nonbipartitetrans}, both
$-\beta^{-1}\dot{I}_Y^{\rm uni}$ and $\dot{J}_Y^{\rm uni}$ contribute to $\dot{F}_Y$. This contrasts with the bipartite case, where the charging is driven solely by the information flow. Nevertheless, the overall scaling behavior is similar: the charging power scales as $O(N^2)$, while the maximum free-energy gain remains $O(1)$.

This enhancement is closely related to supertransfer~\cite{supertransfer}, in which collective interactions allow the effective interaction strength to scale with $N$. This collective enhancement accelerates both the unitary information flow and the unitary energy flow, resulting in an $O(1/N^2)$ charging time.

\section*{Discussion}
Our work establishes a framework that connects the enhancement of quantum information flow to improved performance of protocols for information-mediated energy transfer in interacting open quantum systems. We derived a short-time TUR showing that, in the NESS, the quantum information flow is bounded by the partial entropy production and a partial activity.
A central implication of this bound is that collective system--bath coupling can quantitatively alter the scaling of information-thermodynamic processes. When quantum coherence enhances the partial activity to $O(N^2)$, an $O(N)$ information flow can be sustained at the cost of only $O(1)$ partial entropy production. This scaling goes beyond the scaling imposed by the current-dissipation trade-off relation for non-collective cases.

In the NESS considered here, the information flow is purely dissipative, $\dot{I}_Y=\dot{I}_Y^{\rm diss}$. In this regime, the enhanced information flow serves as a thermodynamic resource for an autonomous Maxwell's demon, enabling an $O(N)$ heat current against a thermal gradient while keeping the total entropy production at $O(1)$. Beyond the NESS, the unitary component $\dot{I}_Y^{\rm uni}$ provides a coherent mechanism for rapidly changing the nonequilibrium free energy of a quantum battery. The collective enhancement accelerates the charging rate to $O(N^2)$ while the stored free energy does not vanish as $N$ increases, yielding a fast charging protocol.

These results show that quantum coherence and collective coupling can enhance quantum information flow, thereby improving information-mediated energy transfer. Our results provide design principles for quantum information-thermodynamic processes by identifying collective dissipative activity and collective interaction strength as resources for controlling heat, work, and information flows in engineered quantum machines. They further suggest natural directions for connecting this mechanism to realistic platforms, where collective states, dissipation, and interactions can be engineered and controlled. Such developments may open a route toward information-powered quantum machines governed by collective quantum effects.

\section{Methods}

\subsection{Details of the scaling analysis of the information flow}
In the NESS, energy conservation gives $\dot Q_X+\sum_i\dot Q_{Y_i}=0$, and hence $\left|\sum_i\dot Q_{Y_i}\right|=|\dot Q_X|$. Since $\dot{I}_{X}=O(N)$, the heat currents must scale at least linearly with $N$, i.e., $\sum_{i}\dot{Q}_{Y_{i}}=O(N)$. This requirement is compatible with an $O(1)$ partial entropy production rate for subsystem $Y$ when the partial activity scales as $A_{Y}=O(N^{2})$. Indeed, applying the short-time TUR~\eqref{TUR} for subsystem $Y$ allows $\sum_{i}\dot{Q}_{Y_{i}}=O(N)$ while keeping $\dot{\sigma}_{Y}=O(1)$.

\subsection{Specific configuration of the autonomous Maxwell's demon model}

To model the collective dissipation, we introduce collective coupling operators for each subsystem, $S_X = \sum_{i,i'}\sigma_{-}^{X_{ii'}}$ and $S_Y = \sum_{j,j'}\sigma_{-}^{Y_{jj'}}$. The corresponding quantum jump operators, expressed in terms of the energy eigenbasis $\ket{\epsilon}$ of $H_{\rm sys}$, are defined as
\begin{align}
    L_{X}^{\omega} &\coloneqq \sum_{\epsilon-\epsilon'=\omega}
    \ket{\epsilon'}\bra{\epsilon'} S_X \otimes \mathbb{I}_Y \ket{\epsilon}\bra{\epsilon}, \\
    L^{\omega}_{Y} &\coloneqq \sum_{\epsilon-\epsilon'=\omega}
    \ket{\epsilon'}\bra{\epsilon'} \mathbb{I}_X \otimes S_Y \ket{\epsilon}\bra{\epsilon}.
\end{align}
In the four-level collective basis $\{\ket{g}, \ket{\psi_1}, 
\ket{\psi_2}, \ket{e}\}$, these jump operators take the explicit form given in Eq.~\eqref{explicitjump}, where the factor of $N$ arises from the collective enhancement, yielding transition rates that scale as $N^2$.

From the TUR, the ratio $\dot{I}_{X}^{2}/\dot{\sigma}_{X}$ is bounded above by $O(N^2)$, so that the individual quantities may scale as $\dot{I}_{X}=O(N^{\alpha})$ and $\dot{\sigma}_{X}=O(N^{2\alpha-2})$, where $\alpha$ is a free parameter. By setting $\mathcal{A}=O(1/N)$, we can fix this free parameter as $\alpha=1$.

\subsection{Decomposition of information flow}
Here, we provide the derivation of the decomposition in Eq.~\eqref{decomposition}. Following Ref.~\cite{QIF}, we obtain
\begin{align}
    \dot I_Y
    &=\Tr[i[H_{\rm sys},\rho]\ln\rho_Y]
    +\Tr[\mathcal{D}_Y[\rho]\ln\rho]
    -\Tr[\mathcal{D}[\rho]\ln\rho_Y].
    \label{decomI}
\end{align}
We further simplify the unitary term as follows. Since $H_{\rm sys}=H_X+H_Y+H_{\rm int}$, the contribution from $H_X$ vanishes because $H_X$ commutes with $\ln\rho_Y$. The contribution from $H_Y$ also vanishes because
\begin{align}
    \Tr[i[H_Y,\rho]\ln\rho_Y] &= 
    \Tr_Y[i[H_Y,\rho_Y]\ln\rho_Y]  = 0.
\end{align}
Thus, only the interaction Hamiltonian contributes to the unitary part. Substituting these identities into Eq.~\eqref{decomI} yields Eq.~\eqref{decomposition}. We note that the steady state $\rho_{\rm ss}$ of the master equation~\eqref{GKSL} satisfies the condition $[H_{\rm sys},\rho_{\rm ss}]=0$ because the steady state of our model is diagonal in the energy eigenbasis. Combined with the expression Eq.~\eqref{decomI}, we find that $\dot{I}_{Y}^{\rm uni}=0$ in the NESS.

\subsection{Bound on unitary information flow}
We derive an upper bound on the unitary information flow:
\begin{equation}\label{unibound}
    |\dot I_Y^{\rm uni}| \leq \sqrt{\mathcal{F}_\rho(H_{\rm{int}})\,
    V_\rho(\ln\rho_Y)},
\end{equation}
where $V_\rho(A)\coloneqq \Tr[\rho A^2]-\Tr[\rho A]^2$ and the quantum Fisher information~\cite{TothApellaniz2014} is $\mathcal{F}_\rho(X)\coloneqq2\sum_{m\neq n}\frac{(p_n-p_m)^2}{p_n+p_m}|\langle n|X|m\rangle|^2$, $\rho=\sum_n p_n\ket n\bra n$. 
Here, $\ln\rho_Y$ is defined on the support of $\rho_Y$ in the collective subspace.
When $V_\rho(\ln\rho_Y)=O(1)$, the scaling of the bound is governed by $\mathcal{F}_\rho(H_{\rm{int}})$. Using the standard bound $\mathcal{F}_\rho(H_{\rm{int}})\leq4V_\rho(H_{\rm int}) \leq \left(\lambda_{\max}(H_{\rm{int}}) - \lambda_{\min}(H_{\rm{int}})\right)^2$~\cite{TothApellaniz2014}, the key quantity is the spectral width of $H_{\rm{int}}$.

\bibliography{ref.bib}

@Article{Vu2026,
  title = {Information-thermodynamic bounds on precision in interacting quantum systems},
  author = {Honma, Ryotaro and Van Vu, Tan},
  journal = {Phys. Rev. A},
  volume = {113},
  issue = {3},
  pages = {032207},
  numpages = {22},
  year = {2026},
  month = {Mar},
  publisher = {American Physical Society},
  doi = {10.1103/f77p-kw54},
  url = {https://link.aps.org/doi/10.1103/f77p-kw54}
}

@Article{TajimaFuno,
  title = {Superconducting-like Heat Current: Effective Cancellation of Current-Dissipation Trade-Off by Quantum Coherence},
  author = {Tajima, Hiroyasu and Funo, Ken},
  journal = {Phys. Rev. Lett.},
  volume = {127},
  issue = {19},
  pages = {190604},
  numpages = {7},
  year = {2021},
  month = {Nov},
  publisher = {American Physical Society},
  doi = {10.1103/PhysRevLett.127.190604},
  url = {https://link.aps.org/doi/10.1103/PhysRevLett.127.190604}
}

@Article{QIF,
  title = {Thermodynamics of Quantum Information Flows},
  author = {Ptaszy\ifmmode \acute{n}\else \'{n}\fi{}ski, Krzysztof and Esposito, Massimiliano},
  journal = {Phys. Rev. Lett.},
  volume = {122},
  issue = {15},
  pages = {150603},
  numpages = {7},
  year = {2019},
  month = {Apr},
  publisher = {American Physical Society},
  doi = {10.1103/PhysRevLett.122.150603},
  url = {https://link.aps.org/doi/10.1103/PhysRevLett.122.150603}
}

@Article{yamauchi,
      title={Thermodynamic speed limit for non-adiabatic work and its classical–quantum decomposition}, 
      author={Aoi Yamauchi and Rihito Nagase and Kaixin Li and Takahiro Sagawa and Ken Funo},
      year={2025},
      journal = {J. Phys. A: Math. Theor.},
    volume = {58},
    pages = {205001},
}

@Article{supertransfer,
      title={Symmetry-enhanced supertransfer of delocalized quantum states}, 
      author={Seth Lloyd and Masoud Mohseni},
      year={2010},
      journal = {New J. Phys.},
    volume = {12},
    pages = {075020},
    doi = {10.1088/1367-2630/12/7/075020}
}

@Article{network,
      title={Network theory of microscopic and macroscopic behavior of master equation systems}, 
      author={J. Schnakenberg},
      year={1976},
      journal = {Rev. Mod. Phys.},
    volume = {48}, 
    pages = {571--585},
    doi = {10.1103/RevModPhys.48.571}
}

@Article{PhysRevX.4.031015,
  title = {Thermodynamics with Continuous Information Flow},
  author = {Horowitz, Jordan M. and Esposito, Massimiliano},
  journal = {Phys. Rev. X},
  volume = {4},
  issue = {3},
  pages = {031015},
  numpages = {11},
  year = {2014},
  month = {Jul},
  publisher = {American Physical Society},
  doi = {10.1103/PhysRevX.4.031015},
  url = {https://link.aps.org/doi/10.1103/PhysRevX.4.031015}
}

@Article{shiraishisaito,
  title = {Universal Trade-Off Relation between Power and Efficiency for Heat Engines},
  author = {Shiraishi, Naoto and Saito, Keiji and Tasaki, Hal},
  journal = {Phys. Rev. Lett.},
  volume = {117},
  issue = {19},
  pages = {190601},
  numpages = {6},
  year = {2016},
  month = {Oct},
  publisher = {American Physical Society},
  doi = {10.1103/PhysRevLett.117.190601},
  url = {https://link.aps.org/doi/10.1103/PhysRevLett.117.190601}
}

@Article{FunoTajima,
  title = {Symmetry Induced Enhancement in Finite-Time Thermodynamic Trade-Off Relations},
  author = {Funo, Ken and Tajima, Hiroyasu},
  journal = {Phys. Rev. Lett.},
  volume = {134},
  issue = {8},
  pages = {080401},
  numpages = {8},
  year = {2025},
  month = {Feb},
  publisher = {American Physical Society},
  doi = {10.1103/PhysRevLett.134.080401},
  url = {https://link.aps.org/doi/10.1103/PhysRevLett.134.080401}
}

@Article{coherentfeedback,
  title = {Quantum effects improve the energy efficiency of feedback control},
  author = {Horowitz, Jordan M. and Jacobs, Kurt},
  journal = {Phys. Rev. E},
  volume = {89},
  issue = {4},
  pages = {042134},
  numpages = {6},
  year = {2014},
  month = {Apr},
  publisher = {American Physical Society},
  doi = {10.1103/PhysRevE.89.042134},
  url = {https://link.aps.org/doi/10.1103/PhysRevE.89.042134}
}

@Article{PhysRevE.100.052137,
  title = {Thermodynamic uncertainty relations including measurement and feedback},
  author = {Potts, Patrick P. and Samuelsson, Peter},
  journal = {Phys. Rev. E},
  volume = {100},
  issue = {5},
  pages = {052137},
  numpages = {8},
  year = {2019},
  month = {Nov},
  publisher = {American Physical Society},
  doi = {10.1103/PhysRevE.100.052137},
  url = {https://link.aps.org/doi/10.1103/PhysRevE.100.052137}
}

@Article{PhysRevLett.114.158101,
  title = {Thermodynamic Uncertainty Relation for Biomolecular Processes},
  author = {Barato, Andre C. and Seifert, Udo},
  journal = {Phys. Rev. Lett.},
  volume = {114},
  issue = {15},
  pages = {158101},
  numpages = {5},
  year = {2015},
  month = {Apr},
  publisher = {American Physical Society},
  doi = {10.1103/PhysRevLett.114.158101},
  url = {https://link.aps.org/doi/10.1103/PhysRevLett.114.158101}
}

@Article{PhysRevLett.116.120601,
  title = {Dissipation Bounds All Steady-State Current Fluctuations},
  author = {Gingrich, Todd R. and Horowitz, Jordan M. and Perunov, Nikolay and England, Jeremy L.},
  journal = {Phys. Rev. Lett.},
  volume = {116},
  issue = {12},
  pages = {120601},
  numpages = {5},
  year = {2016},
  month = {Mar},
  publisher = {American Physical Society},
  doi = {10.1103/PhysRevLett.116.120601},
  url = {https://link.aps.org/doi/10.1103/PhysRevLett.116.120601}
}

@Article{PhysRevLett.128.140602,
  title = {Thermodynamics of Precision in Markovian Open Quantum Dynamics},
  author = {Van Vu, Tan and Saito, Keiji},
  journal = {Phys. Rev. Lett.},
  volume = {128},
  issue = {14},
  pages = {140602},
  numpages = {8},
  year = {2022},
  month = {Apr},
  publisher = {American Physical Society},
  doi = {10.1103/PhysRevLett.128.140602},
  url = {https://link.aps.org/doi/10.1103/PhysRevLett.128.140602}
}

@Article{TURclassicaldemon,
  title = {Universal bounds on the performance of information-thermodynamic engine},
  author = {Tanogami, Tomohiro and Van Vu, Tan and Saito, Keiji},
  journal = {Phys. Rev. Res.},
  volume = {5},
  issue = {4},
  pages = {043280},
  numpages = {24},
  year = {2023},
  month = {Dec},
  publisher = {American Physical Society},
  doi = {10.1103/PhysRevResearch.5.043280},
  url = {https://link.aps.org/doi/10.1103/PhysRevResearch.5.043280}
}

@Article{PRXQuantum.6.010343,
  title = {Fundamental Bounds on Precision and Response for Quantum Trajectory Observables},
  author = {Van Vu, Tan},
  journal = {PRX Quantum},
  volume = {6},
  issue = {1},
  pages = {010343},
  numpages = {22},
  year = {2025},
  month = {Mar},
  publisher = {American Physical Society},
  doi = {10.1103/PRXQuantum.6.010343},
  url = {https://link.aps.org/doi/10.1103/PRXQuantum.6.010343}
}

@Article{NaturephysTUR,
  title = {Thermodynamic uncertainty relations constrain non-equilibrium fluctuations},
  author = {Horowitz, Jordan M. and Gingrich, Todd R.},
  journal = {Nat. Phys.},
  volume = {16},
  pages = {15--20},
  year = {2020},
  doi = {10.1038/s41567-019-0702-6},
  url = {https://doi.org/10.1038/s41567-019-0702-6}
}

@Article{Naturephysdemon,
  title = {Experimental demonstration of information-to-energy conversion and validation of the generalized Jarzynski equality},
  author = {Toyabe, Shoichi and Sagawa, Takahiro and Ueda, Masahito and Muneyuki, Eiro and Sano, Masaki},
  journal = {Nat. Phys.},
  volume = {6},
  pages = {988--992},
  year = {2010},
  doi = {10.1038/nphys1821},
  url = {https://doi.org/10.1038/nphys1821}
}

@Article{NaturephysITD,
  title = {Thermodynamics of information},
  author = {Parrondo, J and Horowitz, J and Sagawa, T},
  journal = {Nat. Phys.},
  volume = {11},
  pages = {131--139},
  year = {2015},
  month = {Feb},
  publisher = {Nature physics},
  doi = {10.1038/nphys3230},
  url = {https://doi.org/10.1038/nphys3230}
}

@Article{PhysRevLett.100.080403,
  title = {Second Law of Thermodynamics with Discrete Quantum Feedback Control},
  author = {Sagawa, Takahiro and Ueda, Masahito},
  journal = {Phys. Rev. Lett.},
  volume = {100},
  issue = {8},
  pages = {080403},
  numpages = {4},
  year = {2008},
  month = {Feb},
  publisher = {American Physical Society},
  doi = {10.1103/PhysRevLett.100.080403},
  url = {https://link.aps.org/doi/10.1103/PhysRevLett.100.080403}
}

@Article{PhysRevLett.104.090602,
  title = {Generalized Jarzynski Equality under Nonequilibrium Feedback Control},
  author = {Sagawa, Takahiro and Ueda, Masahito},
  journal = {Phys. Rev. Lett.},
  volume = {104},
  issue = {9},
  pages = {090602},
  numpages = {4},
  year = {2010},
  month = {Mar},
  publisher = {American Physical Society},
  doi = {10.1103/PhysRevLett.104.090602},
  url = {https://link.aps.org/doi/10.1103/PhysRevLett.104.090602}
}

@Article{PhysRevLett.109.180602,
  title = {Fluctuation Theorem with Information Exchange: Role of Correlations in Stochastic Thermodynamics},
  author = {Sagawa, Takahiro and Ueda, Masahito},
  journal = {Phys. Rev. Lett.},
  volume = {109},
  issue = {18},
  pages = {180602},
  numpages = {5},
  year = {2012},
  month = {Nov},
  publisher = {American Physical Society},
  doi = {10.1103/PhysRevLett.109.180602},
  url = {https://link.aps.org/doi/10.1103/PhysRevLett.109.180602}
}

@Article{PhysRevLett.111.180603,
  title = {Information Thermodynamics on Causal Networks},
  author = {Ito, Sosuke and Sagawa, Takahiro},
  journal = {Phys. Rev. Lett.},
  volume = {111},
  issue = {18},
  pages = {180603},
  numpages = {6},
  year = {2013},
  month = {Oct},
  publisher = {American Physical Society},
  doi = {10.1103/PhysRevLett.111.180603},
  url = {https://link.aps.org/doi/10.1103/PhysRevLett.111.180603}
}

@Article{PhysRevLett.117.240502,
  title = {Experimental Rectification of Entropy Production by Maxwell's Demon in a Quantum System},
  author = {Camati, Patrice A. and Peterson, John P. S. and Batalh\~ao, Tiago B. and Micadei, Kaonan and Souza, Alexandre M. and Sarthour, Roberto S. and Oliveira, Ivan S. and Serra, Roberto M.},
  journal = {Phys. Rev. Lett.},
  volume = {117},
  issue = {24},
  pages = {240502},
  numpages = {6},
  year = {2016},
  month = {Dec},
  publisher = {American Physical Society},
  doi = {10.1103/PhysRevLett.117.240502},
  url = {https://link.aps.org/doi/10.1103/PhysRevLett.117.240502}
}

@Article{PhysRevLett.111.230402,
  title = {Heat Engine Driven by Purely Quantum Information},
  author = {Park, Jung Jun and Kim, Kang-Hwan and Sagawa, Takahiro and Kim, Sang Wook},
  journal = {Phys. Rev. Lett.},
  volume = {111},
  issue = {23},
  pages = {230402},
  numpages = {5},
  year = {2013},
  month = {Dec},
  publisher = {American Physical Society},
  doi = {10.1103/PhysRevLett.111.230402},
  url = {https://link.aps.org/doi/10.1103/PhysRevLett.111.230402}
}

@Article{RevModPhys.96.031001,
  title = {Colloquium: Quantum batteries},
  author = {Campaioli, Francesco and Gherardini, Stefano and Quach, James Q. and Polini, Marco and Andolina, Gian Marcello},
  journal = {Rev. Mod. Phys.},
  volume = {96},
  issue = {3},
  pages = {031001},
  numpages = {30},
  year = {2024},
  month = {Jul},
  publisher = {American Physical Society},
  doi = {10.1103/RevModPhys.96.031001},
  url = {https://link.aps.org/doi/10.1103/RevModPhys.96.031001}
}

@Article{PhysRevLett.134.130401,
  title = {Large Collective Power Enhancement in Dissipative Charging of a Quantum Battery},
  author = {Pokhrel, Sagar and Gea-Banacloche, Julio},
  journal = {Phys. Rev. Lett.},
  volume = {134},
  issue = {13},
  pages = {130401},
  numpages = {6},
  year = {2025},
  month = {Mar},
  publisher = {American Physical Society},
  doi = {10.1103/PhysRevLett.134.130401},
  url = {https://link.aps.org/doi/10.1103/PhysRevLett.134.130401}
}

@Article{Sagawa2012PTP,
  title = {Thermodynamics of Information Processing in Small Systems},
  author = {Sagawa, Takahiro},
  journal = {Prog. Theor. Phys.},
  volume = {127},
  pages = {1--56},
  year = {2012},
  doi = {10.1143/PTP.127.1}
}

@book{BreuerPetruccione,
  title = {The Theory of Open Quantum Systems},
  author = {Breuer, Heinz-Peter and Petruccione, Francesco},
  publisher = {Oxford University Press},
  year = {2002}
}

@Article{Dicke1954,
  title = {Coherence in Spontaneous Radiation Processes},
  author = {Dicke, R. H.},
  journal = {Phys. Rev.},
  volume = {93},
  issue = {1},
  pages = {99--110},
  year = {1954},
  doi = {10.1103/PhysRev.93.99},
  url = {https://link.aps.org/doi/10.1103/PhysRev.93.99}
}

@Article{Universaltradeoffbetween,
  title = {Universal Trade-Off between Power, Efficiency, and Constancy in Steady-State Heat Engines},
  author = {Pietzonka, Patrick and Seifert, Udo},
  journal = {Phys. Rev. Lett.},
  volume = {120},
  issue = {19},
  pages = {190602},
  numpages = {6},
  year = {2018},
  month = {May},
  publisher = {American Physical Society},
  doi = {10.1103/PhysRevLett.120.190602},
  url = {https://link.aps.org/doi/10.1103/PhysRevLett.120.190602}
}

@Article{quantumenhancedheatengine,
  title = {Quantum-Enhanced Heat Engine Based on Superabsorption},
  author = {Kamimura, Shunsuke and Hakoshima, Hideaki and Matsuzaki, Yuichiro and Yoshida, Kyo and Tokura, Yasuhiro},
  journal = {Phys. Rev. Lett.},
  volume = {128},
  issue = {18},
  pages = {180602},
  numpages = {7},
  year = {2022},
  month = {May},
  publisher = {American Physical Society},
  doi = {10.1103/PhysRevLett.128.180602},
  url = {https://link.aps.org/doi/10.1103/PhysRevLett.128.180602}
}

@Article{enhancingquantumbattery,
  title = {Enhancing the Charging Power of Quantum Batteries},
  author = {Campaioli, Francesco and Pollock, Felix A. and Binder, Felix C. and C\'eleri, Lucas and Goold, John and Vinjanampathy, Sai and Modi, Kavan},
  journal = {Phys. Rev. Lett.},
  volume = {118},
  issue = {15},
  pages = {150601},
  numpages = {6},
  year = {2017},
  month = {Apr},
  publisher = {American Physical Society},
  doi = {10.1103/PhysRevLett.118.150601},
  url = {https://link.aps.org/doi/10.1103/PhysRevLett.118.150601}
}

@Article{highpowercollective,
  title = {High-Power Collective Charging of a Solid-State Quantum Battery},
  author = {Ferraro, Dario and Campisi, Michele and Andolina, Gian Marcello and Pellegrini, Vittorio and Polini, Marco},
  journal = {Phys. Rev. Lett.},
  volume = {120},
  issue = {11},
  pages = {117702},
  numpages = {6},
  year = {2018},
  month = {Mar},
  publisher = {American Physical Society},
  doi = {10.1103/PhysRevLett.120.117702},
  url = {https://link.aps.org/doi/10.1103/PhysRevLett.120.117702}
}

@Article{Koski2015Refrigerator,
  title = {On-Chip Maxwell's Demon as an Information-Powered Refrigerator},
  author = {Koski, J. V. and Kutvonen, A. and Khaymovich, I. M. and Ala-Nissila, T. and Pekola, J. P.},
  journal = {Phys. Rev. Lett.},
  volume = {115},
  pages = {260602},
  year = {2015},
  doi = {10.1103/PhysRevLett.115.260602}
}

@misc{HonmaVu2026,
      title={Thermodynamic Uncertainty Relation with Quantum Feedback}, 
      author={Honma, Ryotaro and Van Vu, Tan},
      year={2026},
      eprint={2602.22651},
      archivePrefix={arXiv},
      primaryClass={quant-ph},
      howpublished = {arXiv:2602.22651},
      url={https://arxiv.org/abs/2602.22651}, 
}

@Article{OtsuboItoDechantSagawa2020,
  title = {Estimating entropy production by machine learning of short-time fluctuating currents},
  author = {Otsubo, Shun and Ito, Sosuke and Dechant, Andreas and Sagawa, Takahiro},
  journal = {Phys. Rev. E},
  volume = {101},
  pages = {062106},
  year = {2020},
  doi = {10.1103/PhysRevE.101.062106}
}

@Article{PhysRevX.7.021003,
  title = {Quantum and Information Thermodynamics: A Unifying Framework Based on Repeated Interactions},
  author = {Strasberg, Philipp and Schaller, Gernot and Brandes, Tobias and Esposito, Massimiliano},
  journal = {Phys. Rev. X},
  volume = {7},
  issue = {2},
  pages = {021003},
  numpages = {33},
  year = {2017},
  month = {Apr},
  publisher = {American Physical Society},
  doi = {10.1103/PhysRevX.7.021003},
  url = {https://link.aps.org/doi/10.1103/PhysRevX.7.021003}
}

@Article{PhysRevLett.121.070601,
  title = {Speed Limit for Classical Stochastic Processes},
  author = {Shiraishi, Naoto and Funo, Ken and Saito, Keiji},
  journal = {Phys. Rev. Lett.},
  volume = {121},
  issue = {7},
  pages = {070601},
  numpages = {6},
  year = {2018},
  month = {Aug},
  publisher = {American Physical Society},
  doi = {10.1103/PhysRevLett.121.070601},
  url = {https://link.aps.org/doi/10.1103/PhysRevLett.121.070601}
}

@Article{Funo_2019,
doi = {10.1088/1367-2630/aaf9f5},
url = {https://doi.org/10.1088/1367-2630/aaf9f5},
year = {2019},
month = {jan},
publisher = {IOP Publishing},
volume = {21},
pages = {013006},
author = {Funo, Ken and Shiraishi, Naoto and Saito, Keiji},
title = {Speed limit for open quantum systems},
journal = {New J. Phys.},
}

@Article{Horowitz_2014,
doi = {10.1088/1367-2630/16/12/125007},
url = {https://doi.org/10.1088/1367-2630/16/12/125007},
year = {2014},
month = {dec},
publisher = {IOP Publishing},
volume = {16},
pages = {125007},
author = {Horowitz, Jordan M and Sandberg, Henrik},
title = {Second-law-like inequalities with information and their interpretations},
journal = {New J. Phys.},
}

@Article{TothApellaniz2014,
  title = {Quantum metrology from a quantum information science perspective},
  author = {T{\'o}th, G{\'e}za and Apellaniz, Iagoba},
  journal = {J. Phys. A: Math. Theor.},
  volume = {47},
  pages = {424006},
  year = {2014},
  doi = {10.1088/1751-8113/47/42/424006}
}

@Article{Photocell,
  title = {Efficient Biologically Inspired Photocell Enhanced by Delocalized Quantum States},
  author = {Creatore, C. and Parker, M. A. and Emmott, S. and Chin, A. W.},
  journal = {Phys. Rev. Lett.},
  volume = {111},
  issue = {25},
  pages = {253601},
  numpages = {5},
  year = {2013},
  month = {Dec},
  publisher = {American Physical Society},
  doi = {10.1103/PhysRevLett.111.253601},
  url = {https://link.aps.org/doi/10.1103/PhysRevLett.111.253601}
}

@article{kumasaki2025,
  title = {Thermodynamic approach to quantum cooling limit of continuous Gaussian feedback},
  author = {Kumasaki, Kousuke and Yada, Toshihiro and Funo, Ken and Sagawa, Takahiro},
  journal = {Phys. Rev. Res.},
  volume = {7},
  issue = {4},
  pages = {043147},
  numpages = {15},
  year = {2025},
  month = {Nov},
  publisher = {American Physical Society},
  doi = {10.1103/5cz5-n6jt},
  url = {https://link.aps.org/doi/10.1103/5cz5-n6jt}
}

@article{landauer,
  title = {Collective Advantages in Finite-Time Thermodynamics},
  author = {Rolandi, Alberto and Abiuso, Paolo and Perarnau-Llobet, Mart\'{\i}},
  journal = {Phys. Rev. Lett.},
  volume = {131},
  issue = {21},
  pages = {210401},
  numpages = {7},
  year = {2023},
  month = {Nov},
  publisher = {American Physical Society},
  doi = {10.1103/PhysRevLett.131.210401},
  url = {https://link.aps.org/doi/10.1103/PhysRevLett.131.210401}
}

@misc{tojo2026thermodynamicuncertaintyrelationcontinuous,
      title={Thermodynamic uncertainty relation under continuous measurement and feedback with quantum-classical-transfer entropy}, 
      author={Kaito Tojo and Takahiro Sagawa and Ken Funo},
      year={2026},
      eprint={2602.23110},
      archivePrefix={arXiv},
      primaryClass={cond-mat.stat-mech},
      howpublished = {arXiv:2602.23110},
      url={https://arxiv.org/abs/2602.23110}, 
}

@misc{sekiguchi2024improvementspeedlimitsquantum,
      title={Improvement of Speed Limits: Quantum Effect on the Speed in Open Quantum Systems}, 
      author={Kotaro Sekiguchi and Satoshi Nakajima and Ken Funo and Hiroyasu Tajima},
      year={2024},
      eprint={2410.11604},
      archivePrefix={arXiv},
      primaryClass={quant-ph},
      howpublished = {arXiv:2410.11604},
      url={https://arxiv.org/abs/2410.11604}, 
}

@article{sagawaueda2009,
  title = {Minimal Energy Cost for Thermodynamic Information Processing: Measurement and Information Erasure},
  author = {Sagawa, Takahiro and Ueda, Masahito},
  journal = {Phys. Rev. Lett.},
  volume = {102},
  issue = {25},
  pages = {250602},
  numpages = {4},
  year = {2009},
  month = {Jun},
  publisher = {American Physical Society},
  doi = {10.1103/PhysRevLett.102.250602},
  url = {https://link.aps.org/doi/10.1103/PhysRevLett.102.250602}
}

@article{yada2025PRX,
  title = {Experimentally Probing Entropy Reduction via Iterative Quantum Information Transfer},
  author = {Yada, Toshihiro and Stas, Pieter-Jan and Suleymanzade, Aziza and Knall, Erik N. and Yoshioka, Nobuyuki and Sagawa, Takahiro and Lukin, Mikhail D.},
  journal = {Phys. Rev. X},
  volume = {15},
  issue = {3},
  pages = {031054},
  numpages = {17},
  year = {2025},
  month = {Aug},
  publisher = {American Physical Society},
  doi = {10.1103/5lp2-9sps},
  url = {https://link.aps.org/doi/10.1103/5lp2-9sps},
  cofirst = {yes}
}

@article{Niedenzu_2018,
doi = {10.1088/1367-2630/aaed55},
url = {https://doi.org/10.1088/1367-2630/aaed55},
year = {2018},
month = {nov},
publisher = {IOP Publishing},
volume = {20},
pages = {113038},
author = {Niedenzu, Wolfgang and Kurizki, Gershon},
title = {Cooperative many-body enhancement of quantum thermal machine power},
journal = {New J. Phys.}
}

@article{SuperradiantQuantumHeatEngine,
doi = {10.1038/srep12953},
url = {https://doi.org/10.1038/srep12953},
year = {2015},
month = {aug},
volume = {5},
pages = {12953},
author = {Hardal, Ali {\"U}. C. and M{\"u}stecapl{\i}o{\u{g}}lu, {\"O}zg{\"u}r E.},
title = {Superradiant Quantum Heat Engine},
journal = {Sci. Rep.}
}

@article{autodemonoriginal,
  title = {Thermodynamics of a Physical Model Implementing a Maxwell Demon},
  author = {Strasberg, Philipp and Schaller, Gernot and Brandes, Tobias and Esposito, Massimiliano},
  journal = {Phys. Rev. Lett.},
  volume = {110},
  issue = {4},
  pages = {040601},
  numpages = {5},
  year = {2013},
  month = {Jan},
  publisher = {American Physical Society},
  doi = {10.1103/PhysRevLett.110.040601},
  url = {https://link.aps.org/doi/10.1103/PhysRevLett.110.040601}
}

@article{batteryNFE,
  title = {Fluctuations in Extractable Work Bound the Charging Power of Quantum Batteries},
  author = {Garc\'{\i}a-Pintos, Luis Pedro and Hamma, Alioscia and del Campo, Adolfo},
  journal = {Phys. Rev. Lett.},
  volume = {125},
  issue = {4},
  pages = {040601},
  numpages = {6},
  year = {2020},
  month = {Jul},
  publisher = {American Physical Society},
  doi = {10.1103/PhysRevLett.125.040601},
  url = {https://link.aps.org/doi/10.1103/PhysRevLett.125.040601}
}

@article{Kamin_2023,
doi = {10.1088/1751-8121/acdb11},
url = {https://doi.org/10.1088/1751-8121/acdb11},
year = {2023},
month = {jun},
publisher = {IOP Publishing},
volume = {56},
pages = {275302},
author = {Kamin, F H and Abuali, Z and Ness, H and Salimi, S},
title = {Quantum battery charging by non-equilibrium steady-state currents},
journal = {J. Phys. A: Math. Theor.}
}

@article{Cottet2017PNAS,
  title = {Observing a quantum Maxwell demon at work},
  author = {Cottet, N. and Jezouin, S. and Bretheau, L. and Campagne-Ibarcq, P. and Ficheux, Q. and Anders, J. and Auffeves, A. and Azouit, R. and Rouchon, P. and Huard, B.},
  journal = {Proc. Natl. Acad. Sci. U.S.A.},
  volume = {114},
  pages = {7561--7564},
  year = {2017},
  doi = {10.1073/pnas.1704827114},
  cofirst={yes}
}

@article{Masuyama2018NatCommun,
  title = {Information-to-work conversion by Maxwell's demon in a superconducting circuit quantum electrodynamical system},
  author = {Masuyama, Y. and Funo, K. and Murashita, Y. and Noguchi, A. and Kono, S. and Tabuchi, Y. and Yamazaki, R. and Ueda, M. and Nakamura, Y.},
  journal = {Nat. Commun.},
  volume = {9},
  pages = {1291},
  year = {2018},
  doi = {10.1038/s41467-018-03686-y}
}

@article{Naghiloo2018PRL,
  title = {Information gain and loss for a quantum Maxwell's demon},
  author = {Naghiloo, M. and Alonso, J. J. and Romito, A. and Lutz, E. and Murch, K. W.},
  journal = {Phys. Rev. Lett.},
  volume = {121},
  pages = {030604},
  year = {2018},
  doi = {10.1103/PhysRevLett.121.030604}
}

@article{yada2022,
  title = {Quantum Fluctuation Theorem under Quantum Jumps with Continuous Measurement and Feedback},
  author = {Yada, Toshihiro and Yoshioka, Nobuyuki and Sagawa, Takahiro},
  journal = {Phys. Rev. Lett.},
  volume = {128},
  issue = {17},
  pages = {170601},
  numpages = {6},
  year = {2022},
  month = {Apr},
  publisher = {American Physical Society},
  doi = {10.1103/PhysRevLett.128.170601},
  url = {https://link.aps.org/doi/10.1103/PhysRevLett.128.170601}
}

@article{Vu_2020,
    doi = {10.1088/1751-8121/ab64a4},
    url = {https://doi.org/10.1088/1751-8121/ab64a4},
    year = {2020},
    month = {jan},
    publisher = {IOP Publishing},
    volume = {53},
    pages = {075001},
    author = {Van Vu, Tan and Hasegawa, Yoshihiko},
    title = {Uncertainty relation under information measurement and feedback control},
    journal = {J. Phys. A: Math. Theor.},
}

@Inbook{Funo2018,
author="Funo, Ken
and Ueda, Masahito
and Sagawa, Takahiro",
editor="Binder, Felix
and Correa, Luis A.
and Gogolin, Christian
and Anders, Janet
and Adesso, Gerardo",
title="Quantum Fluctuation Theorems",
bookTitle="Thermodynamics in the Quantum Regime: Fundamental Aspects and New Directions",
year="2018",
publisher="Springer International Publishing",
address="Cham",
pages="249--273",
doi="10.1007/978-3-319-99046-0_10",
url="https://doi.org/10.1007/978-3-319-99046-0_10"
}

@article{PhysRevXVu,
  title = {Thermodynamic Unification of Optimal Transport: Thermodynamic Uncertainty Relation, Minimum Dissipation, and Thermodynamic Speed Limits},
  author = {Van Vu, Tan and Saito, Keiji},
  journal = {Phys. Rev. X},
  volume = {13},
  issue = {1},
  pages = {011013},
  numpages = {45},
  year = {2023},
  month = {Feb},
  publisher = {American Physical Society},
  doi = {10.1103/PhysRevX.13.011013},
  url = {https://link.aps.org/doi/10.1103/PhysRevX.13.011013}
}

@article{yoshimura2026,
  title = {Quasiprobability Thermodynamic Uncertainty Relation},
  author = {Yoshimura, Kohei and Hamazaki, Ryusuke},
  journal = {Phys. Rev. Lett.},
  volume = {136},
  issue = {12},
  pages = {120406},
  numpages = {10},
  year = {2026},
  month = {Mar},
  publisher = {American Physical Society},
  doi = {10.1103/ky8n-9bcy},
  url = {https://link.aps.org/doi/10.1103/ky8n-9bcy}
}

@article{shi2022,
  title = {Entanglement, Coherence, and Extractable Work in Quantum Batteries},
  author = {Shi, Hai-Long and Ding, Shu and Wan, Qing-Kun and Wang, Xiao-Hui and Yang, Wen-Li},
  journal = {Phys. Rev. Lett.},
  volume = {129},
  issue = {13},
  pages = {130602},
  numpages = {6},
  year = {2022},
  month = {Sep},
  publisher = {American Physical Society},
  doi = {10.1103/PhysRevLett.129.130602},
  url = {https://link.aps.org/doi/10.1103/PhysRevLett.129.130602}
}
\section{Acknowledgements}

\subsection{Funding}
This work was supported by the Japan Science and Technology Agency (JST) ERATO Grant JPMJER2302 and Google.org. K.T. acknowledges support from the World-leading Innovative Graduate Study Program for Materials Research, Information, and Technology (MERIT-WINGS), The University of Tokyo. T.S. acknowledges support from the Institute of AI and Beyond, The University of Tokyo. K.F. acknowledges support from the Japan Society for the Promotion of Science (JSPS) KAKENHI Grants JP23K13036, JP24H00831, and JP26H02012.
\clearpage
\onecolumngrid
\setcounter{secnumdepth}{3}

\begin{center}
{\large\bfseries Supplementary Information\par}
\vspace{0.8em}
{Shotaro Oki,$^{1}$ Yuki Kadono,$^{1}$ Kaito Tojo,$^{1}$ Takahiro Sagawa,$^{1,2,3}$ and Ken Funo$^{1}$\par}
\vspace{0.4em}
{\small
$^{1}$Department of Applied Physics, The University of Tokyo, 7-3-1 Hongo, Bunkyo-ku, Tokyo 113-8656, Japan\\
$^{2}$Quantum-Phase Electronics Center (QPEC), The University of Tokyo, 7-3-1 Hongo, Bunkyo-ku, Tokyo 113-8656, Japan\\
$^{3}$Inamori Research Institute for Science (InaRIS), Kyoto-shi, Kyoto 600-8411, Japan\\
\par}
\end{center}
\vspace{1.5em}

%%%%%%%%%% Prefix a "S" to all equations, figures, tables and reset the counter %%%%%%%%%%
\setcounter{equation}{0}
\setcounter{figure}{0}
\setcounter{table}{0}
\setcounter{page}{1}
\makeatletter
\renewcommand{\theequation}{S\arabic{equation}}
\renewcommand{\thefigure}{S\arabic{figure}}
\renewcommand{\bibnumfmt}[1]{[S#1]}
\renewcommand{\citenumfont}[1]{S#1}
%%%%%%%%%% Prefix a "S" to all equations, figures, tables and reset the counter %%%%%%%%%%

\section{Derivation of short-time TUR~\eqref{TUR}}
We first consider the case in which subsystem $X$ is coupled to a single heat bath. The extension to multiple heat baths is straightforward. The partial entropy production rate for subsystem $X$ is
\begin{align}
\dot{\sigma}_X &= \dot S_X - \beta_{X}\dot{Q}_{X} - \dot{I}_X \nonumber\\
&= -\beta_X\Tr[\mathcal{D}_X[\rho]H_{\rm sys}] - \Tr[\mathcal{D}_X[\rho]\ln{\rho}].
\end{align}
In the following derivation, we write $H_{\rm sys}=\sum_E E\Pi_E$ and define $\rho_d\coloneqq\sum_E\Pi_E\rho\Pi_E$. We choose the basis $\{\ket{\epsilon}\}$ such that it diagonalizes each block $\Pi_E\rho\Pi_E$ within the corresponding energy eigenspace. Then $\rho_d=\sum_\epsilon \bra{\epsilon}\rho\ket{\epsilon}\ket{\epsilon}\bra{\epsilon}$, where each $\ket{\epsilon}$ is an eigenstate of $H_{\rm sys}$.
The first term can be evaluated explicitly from the dissipator as
\begin{align}\label{firstterm}
(\text{first term}) &= -\beta_X\sum_\omega\gamma_X(\omega)\Tr\left[\left(L_{X}^{\omega}\rho (L^{\omega}_{X})^{\dagger}-\frac{1}{2}\{(L^{\omega}_{X})^{\dagger}L_{X}^{\omega},\rho\}\right)H_{\rm sys}\right] \nonumber\\&= \beta_X\sum_\omega\gamma_X(\omega)\omega\Tr[L_{X}^{\omega}\rho (L^{\omega}_{X})^{\dagger}] \nonumber\\&= 
\beta_X\sum_\omega\gamma_X(\omega)\omega\Tr[L_{X}^{\omega}\rho_d (L^{\omega}_{X})^{\dagger}] \nonumber\\&=
\beta_X\sum_\omega\gamma_X(\omega)\omega\sum_\epsilon\bra{\epsilon}L_{X}^{\omega}\rho_d (L^{\omega}_{X})^{\dagger}\ket{\epsilon} \nonumber\\&= \beta_X\sum_\omega\gamma_X(\omega)\omega\sum_{\epsilon,\epsilon'}\bra{\epsilon}L_{X}^{\omega}\ket{\epsilon'}\bra{\epsilon'}\rho\ket{\epsilon'}\bra{\epsilon'}(L^{\omega}_{X})^{\dagger}\ket{\epsilon} \nonumber\\&= \beta_X\sum_{\omega,\epsilon,\epsilon'}\omega\bra{\epsilon'}\rho\ket{\epsilon'}W_X^{\epsilon'\rightarrow\epsilon,\omega} \nonumber\\&= \sum_{\omega,\epsilon,\epsilon'}\bra{\epsilon'}\rho\ket{\epsilon'}W_X^{\epsilon'\rightarrow\epsilon,\omega}\ln{\frac{W_X^{\epsilon'\rightarrow\epsilon,\omega}}{W_X^{\epsilon\rightarrow\epsilon',-\omega}}},
\end{align}
where we have defined the transition rate $W_X^{\epsilon'\rightarrow\epsilon,\omega}\coloneqq\gamma_X(\omega)|\bra{\epsilon}L_{X}^{\omega}\ket{\epsilon'}|^2$. In the last equality, we used the identity $L_{X}^{-\omega}=(L^{\omega}_{X})^{\dagger}$ together with the local detailed balance condition $\beta_X\omega=\ln{\frac{\gamma_X(\omega)}{\gamma_X(-\omega)}}=\ln{\frac{W_X^{\epsilon'\rightarrow\epsilon,\omega}}{W_X^{\epsilon\rightarrow\epsilon',-\omega}}}$. The second term is bounded as
\begin{align}\label{secondterm}
(\text{second term}) &= -\Tr[\mathcal{D}_X[\rho]\ln{\rho}] \nonumber\\&\ge -\Tr\left[\mathcal{D}_X[\rho]\sum_\epsilon\ln{\bra{\epsilon}\rho\ket{\epsilon}}\ket{\epsilon}\bra{\epsilon}\right] \nonumber\\&= -\sum_\epsilon\bra{\epsilon}\mathcal{D}_X[\rho]\ket{\epsilon}\ln{\bra{\epsilon}\rho\ket{\epsilon}} \nonumber\\&= -\sum_{\epsilon,\epsilon',\omega}(\bra{\epsilon'}\rho\ket{\epsilon'}W_X^{\epsilon'\rightarrow\epsilon,\omega}-\bra{\epsilon}\rho\ket{\epsilon}W_X^{\epsilon\rightarrow\epsilon',-\omega})\ln{\bra{\epsilon}\rho\ket{\epsilon}} \nonumber\\&= \sum_{\epsilon,\epsilon',\omega}\bra{\epsilon'}\rho\ket{\epsilon'}W_X^{\epsilon'\rightarrow\epsilon,\omega}\ln{\frac{\bra{\epsilon'}\rho\ket{\epsilon'}}{\bra{\epsilon}\rho\ket{\epsilon}}},
\end{align}
where we have used the relation
$-\Tr[\mathcal{D}_X[\rho]\ln\rho]
\ge
-\Tr[\mathcal{D}_X[\rho]\ln\rho_d]$ to obtain the second line [see~\eqref{eq:sigma_coherence_cost} for the derivation].
Combining Eqs.~\eqref{firstterm} and~\eqref{secondterm}, we obtain
\begin{align}
\dot{\sigma}_X &\ge \sum_{\epsilon,\epsilon',\omega}W_X^{\epsilon'\rightarrow\epsilon,\omega}\bra{\epsilon'}\rho\ket{\epsilon'}\ln{\frac{W_X^{\epsilon'\rightarrow\epsilon,\omega}\bra{\epsilon'}\rho\ket{\epsilon'}}{W_X^{\epsilon\rightarrow\epsilon',-\omega}\bra{\epsilon}\rho\ket{\epsilon}}} \nonumber\\&= \sum_{\epsilon,\epsilon',\omega}R_X^{\epsilon'\rightarrow\epsilon,\omega}\ln{\frac{R_X^{\epsilon'\rightarrow\epsilon,\omega}}{R_X^{\epsilon\rightarrow\epsilon',-\omega}}} \ge 0,
\end{align}
where $R_X^{\epsilon'\rightarrow\epsilon,\omega} = W_X^{\epsilon'\rightarrow\epsilon,\omega}\bra{\epsilon'}\rho\ket{\epsilon'}$. The heat current is
\begin{align}
\dot Q_X
&=
\Tr[\mathcal{D}_X[\rho]H_{\rm sys}]
=
-\sum_{\omega,\epsilon,\epsilon'}
\omega R_X^{\epsilon'\to\epsilon,\omega}
\notag\\
&=
-\frac{1}{2}
\sum_{\omega,\epsilon,\epsilon'}
\omega
\left(
R_X^{\epsilon'\to\epsilon,\omega}
-
R_X^{\epsilon\to\epsilon',-\omega}
\right).
\end{align}
Thus,
\begin{align}
(\dot Q_X)^2
&=
\left[
\frac{1}{2}
\sum_{\omega,\epsilon,\epsilon'}
\omega
\left(
R_X^{\epsilon'\to\epsilon,\omega}
-
R_X^{\epsilon\to\epsilon',-\omega}
\right)
\right]^2
\notag\\
&\le
\frac{1}{4}
\left[
\sum_{\omega,\epsilon,\epsilon'}
\frac{
\left(
R_X^{\epsilon'\to\epsilon,\omega}
-
R_X^{\epsilon\to\epsilon',-\omega}
\right)^2
}{
R_X^{\epsilon'\to\epsilon,\omega}
+
R_X^{\epsilon\to\epsilon',-\omega}
}
\right]
\left[
\sum_{\omega,\epsilon,\epsilon'}
\omega^2
\left(
R_X^{\epsilon'\to\epsilon,\omega}
+
R_X^{\epsilon\to\epsilon',-\omega}
\right)
\right]
\notag\\
&\le
\frac{1}{2}\dot\sigma_X A_X.
\end{align}
We therefore obtain the short-time TUR
\begin{equation}\label{HTUR}
(\dot{Q}_X)^2 \le \frac{1}{2}\dot{\sigma}_X A_X.
\end{equation}
The first inequality follows from the Cauchy--Schwarz inequality, and the second from the algebraic bound $\frac{1}{2}(a-b)\ln\frac{a}{b}\ge\frac{(a-b)^2}{a+b}$ for $a,b\ge0$. In the last line, we have identified the partial activity $A_X= \frac{1}{2}\sum_{\omega,\epsilon,\epsilon'}\omega^2(R_X^{\epsilon'\rightarrow\epsilon,\omega}+R_X^{\epsilon\rightarrow\epsilon',-\omega})$. This completes the derivation of the short-time TUR. Equivalently, $A_X$ can be written as
\begin{align}
A_X &= \frac{1}{2}\sum_{\omega,\epsilon,\epsilon'}\omega^2(R_X^{\epsilon'\rightarrow\epsilon,\omega}+R_X^{\epsilon\rightarrow\epsilon',-\omega}) \nonumber\\&= \sum_{\omega,\epsilon,\epsilon'}\omega^2 R_X^{\epsilon'\rightarrow\epsilon,\omega} \nonumber\\&= \sum_{\omega}\omega^2\gamma_X(\omega)\Tr[(L^{\omega}_{X})^{\dagger}L_{X}^{\omega}\rho].
\end{align}
This result agrees with the short-time limit of the relation derived in Ref.~\cite{Vu2026}.

\section{Steady-state analysis of the $N$-fold degenerate two-level model}
In this section, we construct a specific physical setup that realizes the steady-state scaling relations discussed in the main text:
\begin{align}
    &\dot{\sigma}_X=O(1),\qquad \dot{\sigma}_Y=O(1), \\
    &\beta_X\dot Q_X=O(N),\qquad
    \sum_i\beta_{Y_i}\dot Q_{Y_i}=O(N), \\
    &\dot I_X=-\dot I_Y=O(N).
\end{align}

\subsection{Bipartite case with single heat baths for $X$ and $Y$}

We show that the $O(N)$ scaling of heat currents and information flow can be realized under the bipartite condition within Schnakenberg's network theory~\cite{network}. Under the bipartite condition, the population dynamics are closed within the four-dimensional collective eigenbasis. We write the steady state as
\begin{equation}
    \rho = P_g\ket{g}\bra{g} + P_1\ket{\psi_1}\bra{\psi_1} 
    + P_2\ket{\psi_2}\bra{\psi_2} + P_e\ket{e}\bra{e},
\end{equation}
and the population dynamics of the master equation are governed by
\begin{align}\label{mastereqbip}
    \dot{P}_g &= -N^2(\Gamma^{\Delta E_-}_{X\uparrow} + \Gamma^{\Delta E_-}_{Y\uparrow}) P_g 
    + N^2\Gamma^{\Delta E_-}_{X\downarrow} P_1 + N^2\Gamma^{\Delta E_-}_{Y\downarrow} P_2, \\
    \dot{P}_1 &= -N^2(\Gamma^{\Delta E_-}_{X\downarrow} + \Gamma^{\Delta E_+}_{Y\uparrow}) P_1 
    + N^2\Gamma^{\Delta E_-}_{X\uparrow} P_g + N^2\Gamma^{\Delta E_+}_{Y\downarrow} P_e, \\
    \dot{P}_2 &= -N^2(\Gamma^{\Delta E_+}_{X\uparrow} + \Gamma^{\Delta E_-}_{Y\downarrow}) P_2 
    + N^2\Gamma^{\Delta E_+}_{X\downarrow} P_e + N^2\Gamma^{\Delta E_-}_{Y\uparrow} P_g, \\
    \dot{P}_e &= -N^2(\Gamma^{\Delta E_+}_{X\downarrow} + \Gamma^{\Delta E_+}_{Y\downarrow}) P_e 
    + N^2\Gamma^{\Delta E_+}_{X\uparrow} P_2 + N^2\Gamma^{\Delta E_+}_{Y\uparrow} P_1,
\end{align}
where $\Gamma_{X\uparrow}^{\Delta E_\pm}$ and $\Gamma_{X\downarrow}^{\Delta E_\pm}$ denote the upward and downward transition rates induced by bath $X$ at energy gap $\Delta E_\pm$,
\begin{equation}
    \Gamma_{X\uparrow}^{\Delta E_\pm}\coloneqq \gamma_X(-\Delta E_\pm),
    \qquad
    \Gamma_{X\downarrow}^{\Delta E_\pm}\coloneqq \gamma_X(\Delta E_\pm),
\end{equation}
with analogous definitions for bath $Y$. A key feature of the bipartite structure is that each transition is assigned to a single subsystem: the transitions $\ket{g}\leftrightarrow\ket{\psi_1}$ and $\ket{\psi_2}\leftrightarrow\ket{e}$ are driven by bath $X$, whereas $\ket{g}\leftrightarrow\ket{\psi_2}$ and $\ket{\psi_1}\leftrightarrow\ket{e}$ are driven by bath $Y$. These four transitions form a single thermodynamic cycle. According to Schnakenberg's network theory~\cite{network}, the stationary probability of each state can be written in terms of maximal tree weights. Let $S_i$ denote the sum of the products of transition rates over all maximal trees directed toward state $i$, and let $S=\sum_i S_i$. Then, the steady-state probability distribution is given by $P_{i}^{\rm ss}=S_i/S$. We define the forward cycle as $g \to \psi_1 \to e \to \psi_2 \to g$, and the reverse cycle as $g \to \psi_2 \to e \to \psi_1 \to g$. The product of transition rates along the forward cycle is
\begin{equation}
\Pi_+ = N^8 \Gamma_{X\uparrow}^{\Delta E_-} \Gamma^{\Delta E_+}_{Y\uparrow} \Gamma_{X\downarrow}^{\Delta E_+} \Gamma^{\Delta E_-}_{Y\downarrow}.
\end{equation}
Similarly, the product of the transition rates along the reverse cycle is
\begin{equation}
\Pi_- = N^8 \Gamma^{\Delta E_-}_{Y\uparrow }\Gamma_{X\uparrow}^{\Delta E_+} \Gamma^{\Delta E_+}_{Y\downarrow }\Gamma_{X\downarrow}^{\Delta E_-}.
\end{equation}
The prefactor $N^8$ appears because each of the four transition rates in the cycle is enhanced by a factor of $N^2$ by the collective jump operators. The steady-state probability current from state $j$ to state $i$ is $\overline{J}_{ij} = W_{j \to i} P^{\rm ss}_j - W_{i \to j} P^{\rm ss}_i$, where $W$ denotes the transition rate. For a single-cycle network, this steady-state flux is uniform on all transition edges. Denoting the uniform cycle flux by $\overline{J}$, we have
\begin{equation}
\overline{J} = \frac{\Pi_+ - \Pi_-}{S}.
\end{equation}
The thermodynamic affinity of this cycle is
\begin{equation}
\mathcal{A} = \ln\frac{\Pi_+}{\Pi_-}.
\end{equation}
Using local detailed balance and $\Delta E_+-\Delta E_-=4\lambda$, the affinity becomes
\begin{equation}
    \mathcal{A}=4\lambda(\beta_X-\beta_Y).
\end{equation}
To analyze the near-equilibrium regime, we take the affinity to scale as
\begin{equation}
\mathcal{A} = O\left(\frac{1}{N}\right).
\end{equation}
Under this condition, the uniform cycle current is $\overline{J} = \left(\frac{\Pi_+}{\Pi_-} - 1\right)\frac{\Pi_-}{S} \approx \mathcal{A}\frac{\Pi_-}{S}$. Because a maximal tree in a four-state network contains three edges, the denominator scales as $S = O((N^2)^3) = O(N^6)$. Since $\Pi_- = O(N^8)$, the uniform cycle current scales as $\overline{J} = O(1/N) \times O(N^8)/O(N^6) = O(N)$. The steady-state heat currents from baths $X$ and $Y$ into the system are therefore determined by the net energy exchange along the cycle:
\begin{align}
    \dot Q_X &= -4\lambda \overline{J}, \\
    \dot Q_Y &=  4\lambda \overline{J}.
\end{align}
Thus, the heat currents are $\dot Q_X = O(N)$ and $\dot Q_Y = O(N)$. The partial entropy production rate $\dot\sigma_X$ can be evaluated from the edge currents associated with the $X$-induced transitions:
\begin{align}
\dot\sigma_X &= N^2(\Gamma_{X\uparrow}^{\Delta E_-}P_g-\Gamma_{X\downarrow}^{\Delta E_-}P_1)\ln\frac{\Gamma_{X\uparrow}^{\Delta E_-}P_g}{\Gamma_{X\downarrow}^{\Delta E_-}P_1} + N^2(\Gamma_{X\uparrow}^{\Delta E_+}P_2-\Gamma_{X\downarrow}^{\Delta E_+}P_e)\ln\frac{\Gamma_{X\uparrow}^{\Delta E_+}P_2}{\Gamma_{X\downarrow}^{\Delta E_+}P_e} \nonumber \\&= \overline{J}\ln\left(1+{\frac{\overline{J}}{N^2\Gamma_{X\downarrow}^{\Delta E_-}P_1}}\right) - \overline{J}\ln\left(1-\frac{\overline{J}}{N^2\Gamma_{X\downarrow}^{\Delta E_+}P_e}\right) \nonumber \\&= O(1)
\end{align}
An analogous argument gives $\dot\sigma_Y=O(1)$. Since the flux $\overline{J}$ is $O(N)$ whereas the denominator $N^2\Gamma_\downarrow P$ is $O(N^2)$, the logarithmic terms scale as $O(1/N)$, and the product remains $O(1)$.

\subsection{Autonomous Maxwell's demon}

We now extend the single-bath setup above to the autonomous Maxwell's demon configuration. Subsystem $Y$ is coupled to two heat baths, labeled $Y_1$ and $Y_2$, whose inverse temperatures satisfy
\begin{equation}
    \beta_X>\beta_{Y_2}>\beta_{Y_1},
\end{equation}
so that $X$ is coupled to the coldest bath, $Y_1$ to the hottest bath, and $Y_2$ to an intermediate-temperature bath. To make the bath couplings energy selective, we assume that the spectral density of bath $Y_1$ is sharply peaked at the smaller gap $\Delta E_-$. Thus, bath $Y_1$ effectively does not induce transitions at the larger gap $\Delta E_+$ as $\Gamma^{\Delta E_{+}}_{Y_{1}\uparrow}\simeq 0,\Gamma^{\Delta E_{+}}_{Y_{1}\downarrow}\simeq 0$. Similarly, we assume that the spectral density of bath $Y_2$ is sharply peaked at the larger gap $\Delta E_+$, then $\Gamma^{\Delta E_-}_{Y_2\uparrow}\simeq0,\Gamma^{\Delta E_-}_{Y_2\downarrow}\simeq0$. Under this energy-selective coupling configuration, the population dynamics are governed by
\begin{align}\label{mastereqdemon}
    \dot{P}_g &= -N^2(\Gamma^{\Delta E_-}_{X\uparrow} + \Gamma^{\Delta E_-}_{Y_1\uparrow}) P_g 
    + N^2\Gamma^{\Delta E_-}_{X\downarrow} P_1 + N^2\Gamma^{\Delta E_-}_{Y_1\downarrow} P_2, \\
    \dot{P}_1 &= -N^2(\Gamma^{\Delta E_-}_{X\downarrow} + \Gamma^{\Delta E_+}_{Y_2\uparrow}) P_1 
    + N^2\Gamma^{\Delta E_-}_{X\uparrow} P_g + N^2\Gamma^{\Delta E_+}_{Y_2\downarrow} P_e, \\
    \dot{P}_2 &= -N^2(\Gamma^{\Delta E_+}_{X\uparrow} + \Gamma^{\Delta E_-}_{Y_1\downarrow}) P_2 
    + N^2\Gamma^{\Delta E_+}_{X\downarrow} P_e + N^2\Gamma^{\Delta E_-}_{Y_1\uparrow} P_g, \\
    \dot{P}_e &= -N^2(\Gamma^{\Delta E_+}_{X\downarrow} + \Gamma^{\Delta E_+}_{Y_2\downarrow}) P_e 
    + N^2\Gamma^{\Delta E_+}_{X\uparrow} P_2 + N^2\Gamma^{\Delta E_+}_{Y_2\uparrow} P_1.
\end{align}
The thermodynamic affinity that determines the direction of the cycle is
\begin{equation}
    \mathcal{A}
    =
    \beta_X(\Delta E_+-\Delta E_-)
    +
    \beta_{Y_1}\Delta E_-
    -
    \beta_{Y_2}\Delta E_+
    =
    4\lambda\beta_X
    +
    (\omega-2\lambda)\beta_{Y_1}
    -
    (\omega+2\lambda)\beta_{Y_2}.
\end{equation}
To operate the system as a near-equilibrium autonomous demon, we tune the parameters so that the affinity is strictly positive but scales inversely with $N$:
\begin{equation}
\mathcal{A} = O\left(\frac{1}{N}\right) > 0.
\end{equation}
Applying the same network-theory analysis as in the previous section, this $O(1/N)$ affinity drives a flux $\overline{J} = O(N)$. Since the cycle runs in the forward direction ($\overline{J} > 0$), the steady-state heat currents exchanged with the baths are
\begin{align}
    \dot Q_X &= -4\lambda\overline{J}=O(N)<0, \\
    \dot Q_{Y_1} &= -(\omega-2\lambda)\overline{J}=O(N)<0, \\
    \dot Q_{Y_2} &=  (\omega+2\lambda)\overline{J}=O(N)>0.
\end{align}
Thus, the model pumps an $O(N)$ heat current from the colder bath $Y_2$ to the hotter bath $Y_1$. Together with the entropy-production estimate above, this shows that the demon generates an $O(N)$ reverse heat current while keeping the entropy production at $O(1)$.

\subsection{Numerical analysis of the autonomous Maxwell's demon model}

All numerical results were obtained by solving the master equation, Eqs. (S28)–(S31), projected onto the 
four-level collective basis $\{\ket{g}, \ket{\psi_1}, \ket{\psi_2}, 
\ket{e}\}$. On this basis, the master equation takes the matrix form
\begin{equation}
    \dot{\mathbf{P}}(t) = \Gamma\, \mathbf{P}(t),
\end{equation}
where $\mathbf{P}(t) = (P_g, P_{\psi_1}, P_{\psi_2}, P_e)^{\top}$ 
is the vector of populations and $\Gamma$ is the transition rate 
matrix. The upward transition rates 
are set to $\Gamma^{\Delta E_\pm}_{k\uparrow} = \gamma_0 = 1.0$, and the downward rates are determined by the local detailed balance condition, $\Gamma^{\Delta E_\pm}_{k\downarrow} = \gamma_0\,e^{\beta_k \Delta E_\pm}$, where $k\in\{X,Y_1,Y_2\}$, with the $N^2$ prefactor arising from collective coupling. 

The stationary distribution $\mathbf{P}_{\rm ss}$ was obtained by solving $\Gamma\mathbf{P}_{\rm ss}=0$ with the normalization condition $\sum_\epsilon P_\epsilon=1$ using \texttt{numpy.linalg.solve}.
The parameters were set as follows. The excitation 
energy was fixed at $\omega = 1.0$, and the interaction strength was set to $\lambda=0.1$, giving $\Delta E_\pm=\omega\pm2\lambda=\omega\pm0.2$. The inverse temperatures 
were parameterized as $\beta_X = \beta_0 + \delta$, $\beta_{Y_1} 
= \beta_0 - 0.1\delta$, and $\beta_{Y_2} = \beta_0$, with $\beta_0 
= 1.0$ and $\delta = 0.1/N$, ensuring $\mathcal{A} = O(1/N)$ in the near-equilibrium regime.

\subsection{Non-bipartite case with single heat baths for $X$ and $Y$}

We next consider the case in which subsystems $X$ and $Y$ are each coupled to a single independent heat bath. In this setting, the same scaling can also be established. We assume that the density matrix is diagonal in the energy eigenbasis:
\begin{equation}
\rho(t)
=
P_+(t)\ket{+}\bra{+}
+
P_-(t)\ket{-}\bra{-}
+
P_g(t)\ket{g}\bra{g}
+
P_e(t)\ket{e}\bra{e},
\end{equation}
where $\ket{\pm}=\frac{1}{\sqrt{2}}(\ket{\psi_1}\pm\ket{\psi_2})$. In this configuration, the jump operator for subsystem $X$ associated with the energy gap $\omega_\pm\coloneqq\omega\pm gN^2>0$ is
\begin{align}
L_{X}^{\omega_+} &= \bra{g}\left(\sum_{i,i'}\sigma_-^{X_{ii'}}\otimes \mathbb{I}_Y\right)\ket{+}\ket{g}\bra{+} + \bra{-}\left(\sum_{i,i'}\sigma_-^{X_{ii'}}\otimes \mathbb{I}_Y\right)\ket{e}\ket{-}\bra{e} \nonumber\\&= \left(\frac{1}{N}\sum_{i,j}\bra{g,i_X}\bra{g,j_Y}\right)\left(\sum_{i,i'}\sigma_-^{X_{ii'}}\otimes \mathbb{I}_Y\right)\frac{1}{\sqrt{2}}\left(\frac{1}{N}\sum_{i,j}\ket{e,i_X}\ket{g,j_Y}\right)\ket{g}\bra{+} \nonumber\\&\qquad- \frac{1}{\sqrt{2}}\left(\frac{1}{N}\sum_{i,j}\bra{g,i_X}\bra{e,j_Y}\right)\left(\sum_{i,i'}\sigma_-^{X_{ii'}}\otimes \mathbb{I}_Y\right)\left(\frac{1}{N}\sum_{i,j}\ket{e,i_X}\ket{e,j_Y}\right)\ket{-}\bra{e} \nonumber\\
&= \frac{N}{\sqrt{2}}\ket{g}\bra{+} - \frac{N}{\sqrt{2}}\ket{-}\bra{e}
\end{align}
Similarly, in the basis $\{\ket{g},\ket{+},\ket{-},\ket{e}\}$, the jump operators are
\begin{align}
    L_{X}^{\omega_+} &= \frac{N}{\sqrt{2}}
    \left(\ket{g}\bra{+}-\ket{-}\bra{e}\right), \qquad
    L_{X}^{\omega_-} = \frac{N}{\sqrt{2}}
    \left(\ket{g}\bra{-}+\ket{+}\bra{e}\right), \\
    L_{Y}^{\omega_+} &= \frac{N}{\sqrt{2}}
    \left(\ket{g}\bra{+}+\ket{-}\bra{e}\right), \qquad
    L_{Y}^{\omega_-} = \frac{N}{\sqrt{2}}
    \left(-\ket{g}\bra{-}+\ket{+}\bra{e}\right).
\end{align}
Because the jump operators are collective, their transition matrix elements are of order $N$. For the downward jumps, we have
\begin{align}
    \Tr\!\left[
    (L_X^{\omega_+})^\dagger L_X^{\omega_+}
    \ket{+}\bra{+}
    \right]
    &=
    \Tr\!\left[
    (L_X^{\omega_+})^\dagger L_X^{\omega_+}
    \ket{e}\bra{e}
    \right]
    =
    \frac{N^2}{2}, \\
    \Tr\!\left[
    (L_X^{\omega_-})^\dagger L_X^{\omega_-}
    \ket{-}\bra{-}
    \right]
    &=
    \Tr\!\left[
    (L_X^{\omega_-})^\dagger L_X^{\omega_-}
    \ket{e}\bra{e}
    \right]
    =
    \frac{N^2}{2}.
\end{align}
Equivalently, the corresponding upward transitions generated by $(L_X^{\omega_\pm})^\dagger$ have the same $N^2/2$ enhancement. Thus, all transition rates induced by these collective jumps scale as $O(N^2)$. The same calculation applies to bath $Y$ after replacing $X \to Y$. The resulting population master equation is
\begin{align}\dot P_g(t) &= -\frac{N^2}{2}(\Gamma_{X\uparrow}^{\omega_+}+\Gamma_{X\uparrow}^{\omega_-}+\Gamma_{Y\uparrow}^{\omega_+}+\Gamma_{Y\uparrow}^{\omega_-}) P_g(t) + \frac{N^2}{2}(\Gamma_{X\downarrow}^{\omega_+}+\Gamma_{Y\downarrow}^{\omega_+}) P_+(t) + \frac{N^2}{2}(\Gamma_{X\downarrow}^{\omega_-}+\Gamma_{Y\downarrow}^{\omega_-}) P_-(t),\\
\dot P_+(t) &= \frac{N^2}{2}(\Gamma_{X\uparrow}^{\omega_+}+\Gamma_{Y\uparrow}^{\omega_+}) P_g(t) - \frac{N^2}{2}(\Gamma_{X\downarrow}^{\omega_+}+\Gamma_{Y\downarrow}^{\omega_+}+\Gamma_{X\uparrow}^{\omega_-}+\Gamma_{Y\uparrow}^{\omega_-}) P_+(t) + \frac{N^2}{2}(\Gamma_{X\downarrow}^{\omega_-}+\Gamma_{Y\downarrow}^{\omega_-})P_e(t),\\
\dot P_-(t) &= \frac{N^2}{2}(\Gamma_{X\uparrow}^{\omega_-}+\Gamma_{Y\uparrow}^{\omega_-}) P_g(t) - \frac{N^2}{2}(\Gamma_{X\downarrow}^{\omega_-}+\Gamma_{Y\downarrow}^{\omega_-}+\Gamma_{X\uparrow}^{\omega_+}+\Gamma_{Y\uparrow}^{\omega_+}) P_-(t) + \frac{N^2}{2}(\Gamma_{X\downarrow}^{\omega_+}+\Gamma_{Y\downarrow}^{\omega_+})P_e(t),\\
\dot P_e(t) &= -\frac{N^2}{2}(\Gamma_{X\downarrow}^{\omega_-}+\Gamma_{Y\downarrow}^{\omega_-}+\Gamma_{X\downarrow}^{\omega_+}+\Gamma_{Y\downarrow}^{\omega_+})P_e(t) + \frac{N^2}{2}(\Gamma_{X\uparrow}^{\omega_-}+\Gamma_{Y\uparrow}^{\omega_-})P_+(t) + \frac{N^2}{2}(\Gamma_{X\uparrow}^{\omega_+}+\Gamma_{Y\uparrow}^{\omega_+})P_-(t).
\end{align}
Here, $\Gamma_{X\uparrow}^{\omega_+}\coloneqq\gamma_X(-\omega_+)$ and $\Gamma_{X\downarrow}^{\omega_+}\coloneqq\gamma_X(\omega_+)$ denote the transition rates associated with the larger energy gap $\omega_+$, while $\Gamma_{X\uparrow}^{\omega_-}\coloneqq\gamma_X(-\omega_-)$ and $\Gamma_{X\downarrow}^{\omega_-}\coloneqq\gamma_X(\omega_-)$ correspond to the smaller gap $\omega_-$. The same notation applies to bath $Y$. These equations can be written compactly in matrix form as
\begin{equation}
    \dot {\mathbf{P}}(t)=(M_X+M_Y)\mathbf{P}(t),
\end{equation}
where the population vector is
\begin{equation}
\mathbf{P}(t) = \begin{pmatrix}
P_g(t) \\P_+(t) \\P_-(t) \\P_e(t)
\end{pmatrix}
\end{equation}
and $M_X$ is the transition matrix generated by the dissipator of subsystem $X$; $M_Y$ is defined analogously.
We impose the local detailed balance conditions for each bath:
\begin{equation}
    \frac{\Gamma_{X\downarrow}^{\omega_\pm}}
    {\Gamma_{X\uparrow}^{\omega_\pm}}
    =
    e^{\beta_X\omega_\pm},
    \qquad
    \frac{\Gamma_{Y\downarrow}^{\omega_\pm}}
    {\Gamma_{Y\uparrow}^{\omega_\pm}}
    =
    e^{\beta_Y\omega_\pm}.
\end{equation}
Next, we introduce a reference state $P^{(X)}$, defined as the steady state that would be reached if only bath $X$ were present. This state satisfies local detailed balance with respect to bath $X$:
\begin{equation}
\Gamma_{X\uparrow}^{\omega_+}P_g^{(X)} = \Gamma_{X\downarrow}^{\omega_+}P_+^{(X)}
\end{equation}
\begin{equation}
\Gamma_{X\uparrow}^{\omega_-}P_g^{(X)} = \Gamma_{X\downarrow}^{\omega_-}P_-^{(X)}
\end{equation}
\begin{equation}\Gamma_{X\uparrow}^{\omega_+}P_-^{(X)} = \Gamma_{X\downarrow}^{\omega_+}P_e^{(X)}
\end{equation}
\begin{equation}
\Gamma_{X\uparrow}^{\omega_-}P_+^{(X)} = \Gamma_{X\downarrow}^{\omega_-}P_e^{(X)}
\end{equation}
If each bare transition rate $\Gamma$ is $O(1)$, the components of $P^{(X)}$ are also $O(1)$. By definition, this reference state satisfies
\begin{equation}
M_X P^{(X)} = 0
\end{equation}
We write the actual nonequilibrium steady state as $P^{\rm ss} = P^{(X)} + \delta P$, treating $\delta P$ as a small deviation from the reference state. The steady-state condition $(M_X + M_Y) P^{\rm ss} = 0$ gives
\begin{equation}
M_Y P^{(X)} + (M_X + M_Y) \delta P = 0
\end{equation}
To analyze the near-equilibrium driving regime, we introduce a small temperature difference $\beta_X - \beta_Y > 0$ through
\begin{equation}
(\beta_X - \beta_Y)\omega = \ln\frac{1+a_N}{1-a_N}
\end{equation}
where $a_N = O(1/N)$, and we assume that the exchange interaction strength scales as $g = O(1/N^2)$. We also assume that the downward transition rates are symmetric for the two baths:
\begin{equation}
\Gamma_{X\downarrow}^{\omega_+} = \Gamma_{Y\downarrow}^{\omega_+}
\end{equation}
\begin{equation}
\Gamma_{X\downarrow}^{\omega_-} = \Gamma_{Y\downarrow}^{\omega_-}
\end{equation}
Using local detailed balance and the assumptions above, the upward rates for the two baths differ by $O(1/N)$:
\begin{equation}
\Gamma_{Y\uparrow}^{\omega_\pm}
-
\Gamma_{X\uparrow}^{\omega_\pm}
=
O(1/N).
\end{equation}
By adding and subtracting $M_XP^{(X)}=0$, we obtain
\begin{equation}
    (M_Y-M_X)P^{(X)}+(M_X+M_Y)\delta P=0 .
\end{equation}
The correction satisfies $\sum_i\delta P_i=0$. On this normalization-preserving subspace, $M_X+M_Y=N^2K$ with $K^{-1}=O(1)$ for the irreducible transition network considered here, while $M_Y-M_X=O(N)$ due to the $O(1/N)$ temperature gradient. Hence,
\begin{equation}
    \delta P=O(1/N).
\end{equation}
Now we evaluate the thermodynamic quantities. The heat current from bath $X$ into the system is given by:
\begin{multline}
\dot{Q}_X(t) = \frac{N^2}{2}\omega_+(\Gamma_{X\uparrow}^{\omega_+}P_g(t)-\Gamma_{X\downarrow}^{\omega_+}P_+(t)) + \frac{N^2}{2}\omega_-(\Gamma_{X\uparrow}^{\omega_-}P_g(t)-\Gamma_{X\downarrow}^{\omega_-}P_-(t)) \\+ \frac{N^2}{2}\omega_-(\Gamma_{X\uparrow}^{\omega_-}P_+(t)-\Gamma_{X\downarrow}^{\omega_-}P_e(t)) + \frac{N^2}{2}\omega_+(\Gamma_{X\uparrow}^{\omega_+}P_-(t)-\Gamma_{X\downarrow}^{\omega_+}P_e(t))
\end{multline}
The partial entropy production rate for subsystem $X$ is
\begin{multline}
\dot\sigma_X = \frac{N^2}{2}(\Gamma_{X\uparrow}^{\omega_+}P_g(t)-\Gamma_{X\downarrow}^{\omega_+}P_+(t))\ln\frac{\Gamma_{X\uparrow}^{\omega_+}P_g(t)}{\Gamma_{X\downarrow}^{\omega_+}P_+(t)} + \frac{N^2}{2}(\Gamma_{X\uparrow}^{\omega_-}P_g(t)-\Gamma_{X\downarrow}^{\omega_-}P_-(t))\ln\frac{\Gamma_{X\uparrow}^{\omega_-}P_g(t)}{\Gamma_{X\downarrow}^{\omega_-}P_-(t)} \\+ \frac{N^2}{2}(\Gamma_{X\uparrow}^{\omega_-}P_+(t)-\Gamma_{X\downarrow}^{\omega_-}P_e(t))\ln\frac{\Gamma_{X\uparrow}^{\omega_-}P_+(t)}{\Gamma_{X\downarrow}^{\omega_-}P_e(t)} + \frac{N^2}{2}(\Gamma_{X\uparrow}^{\omega_+}P_-(t)-\Gamma_{X\downarrow}^{\omega_+}P_e(t))\ln\frac{\Gamma_{X\uparrow}^{\omega_+}P_-(t)}{\Gamma_{X\downarrow}^{\omega_+}P_e(t)}
\end{multline}
In the steady state, the transition fluxes scale as
\begin{equation}
\Gamma_{X\uparrow}^{\omega_+}P_g^{\rm ss} - \Gamma_{X\downarrow}^{\omega_+}P_+^{\rm ss} = \Gamma_{X\uparrow}^{\omega_+}\delta P_g^{\rm ss} - \Gamma_{X\downarrow}^{\omega_+}\delta P_+^{\rm ss} = O\left(\frac{1}{N}\right)
\end{equation} Since this $O(1/N)$ scaling holds for all other transition edges, multiplying by the $O(N^2)$ prefactor yields an $O(N)$ heat current:
\begin{equation}
\dot{Q}_X = O(N)
\end{equation}
Furthermore, for the logarithmic affinity terms, we have:
\begin{equation}
\ln\frac{\Gamma_{X\uparrow}^{\omega_+}P_g^{\rm ss}}{\Gamma_{X\downarrow}^{\omega_+}P_+^{\rm ss}} \approx \ln\left(1+O\left(\frac{1}{N}\right)\right) = O\left(\frac{1}{N}\right)
\end{equation}
\begin{equation}
\ln\frac{\Gamma_{X\uparrow}^{\omega_-}P_g^{\rm ss}}{\Gamma_{X\downarrow}^{\omega_-}P_-^{\rm ss}} \approx \ln\left(1+O\left(\frac{1}{N}\right)\right) = O\left(\frac{1}{N}\right)
\end{equation}
Using these approximations for all transitions, each contribution to $\dot\sigma_X$ consists of an $O(1/N)$ flux multiplied by an $O(1/N)$ affinity and the $O(N^2)$ collective prefactor. Hence, $\dot\sigma_X$ remains $O(1)$.
By definition, the information flow in the steady state is $\dot I_X = -\dot \sigma_X - \beta_X\dot Q_X$. From the scalings above, we obtain
\begin{equation}
\dot I_X = O(N)
\end{equation}
The partial activity likewise scales as
\begin{equation}
A_X = O(N^2)
\end{equation}
These results explicitly confirm the anomalous scaling predicted by the information TUR. Because global energy conservation in the steady state gives $\dot Q_X=-\dot Q_Y$, we also have $\dot Q_Y = O(N)$. Since the total information flow vanishes in the steady state,
\begin{equation}
\dot I_X + \dot I_Y = 0
\end{equation}
we obtain $\dot I_Y = O(N)$. The total entropy production then scales as
\begin{equation}
\dot\sigma_X+\dot\sigma_Y
=
-\beta_X\dot Q_X-\beta_Y\dot Q_Y
=
-(\beta_X-\beta_Y)\dot Q_X
=
O(1),
\end{equation}
because $\beta_X-\beta_Y=O(1/N)$ and $\dot Q_X=O(N)$. Since $\dot\sigma_X=O(1)$, it follows that $\dot\sigma_Y=O(1)$. Thus, in the single-bath configuration, the non-bipartite model realizes the same scaling as the bipartite model: $O(N)$ heat currents and information flow with $O(1)$ entropy production.

\subsection{Comparison of bipartite and non-bipartite cases in steady state dynamics} 
In the non-bipartite model, the collective ground state $\ket{g}$ and fully excited state $\ket{e}$ remain eigenstates of $H_{\rm sys}$ with eigenvalues $0$ and $2\omega$, while the exchange interaction hybridizes the single-excitation states into $\ket{\pm} = \frac{1}{\sqrt{2}}(\ket{\psi_1} \pm \ket{\psi_2})$ with eigenvalues $\omega \pm gN^2$.

When subsystem $Y$ is coupled to only a single bath, the bipartite and non-bipartite models exhibit the same essential scaling behavior: they support $O(N)$ heat currents and information flow while maintaining an $O(1)$ partial entropy production rate in the near-equilibrium regime. 

\begin{figure}
    \centering
    \includegraphics[width=0.6\linewidth]{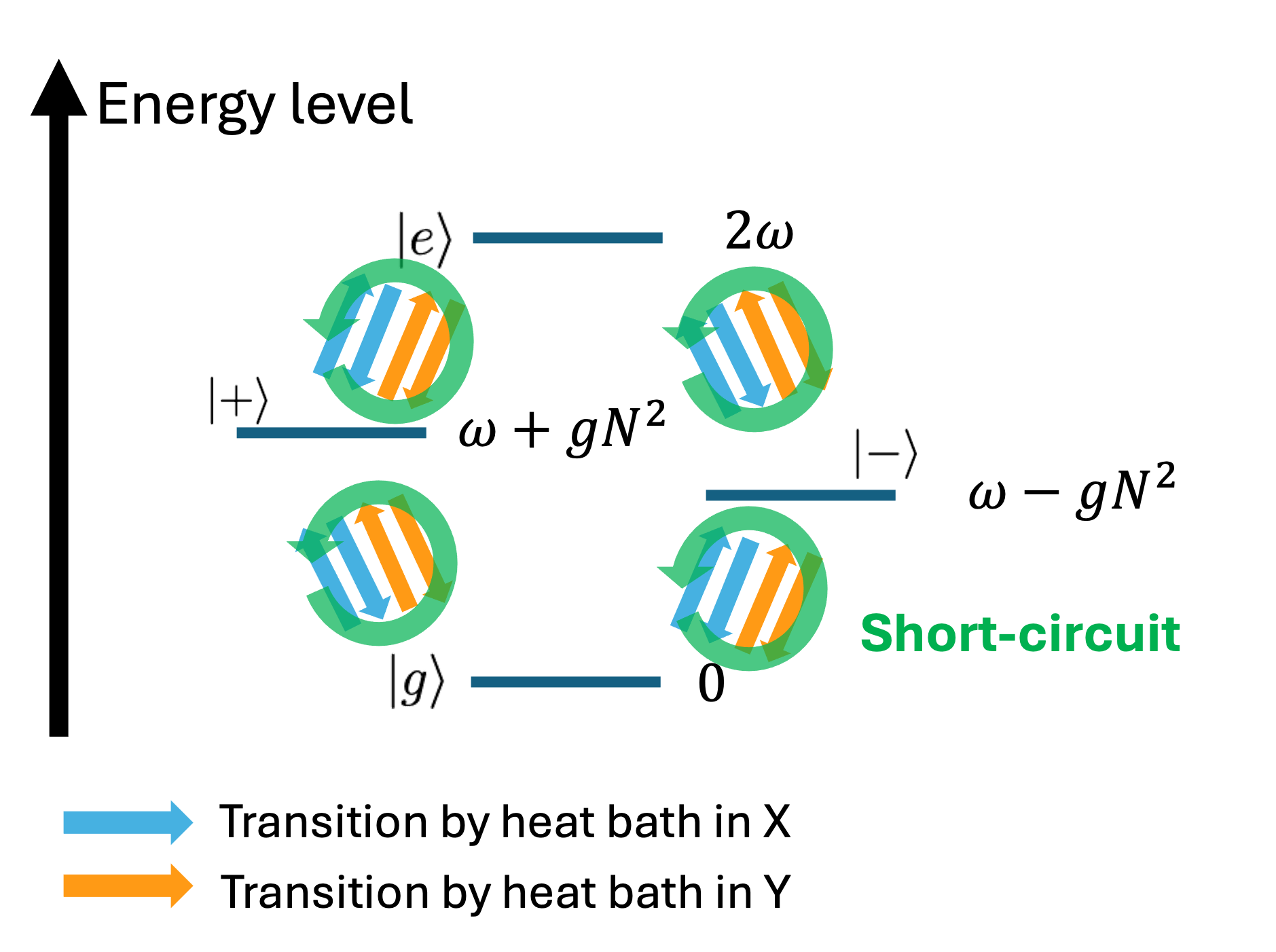}
    \caption{Energy-level diagram of the non-bipartite model. The exchange interaction hybridizes the single-excitation states into $\ket{+}$ and $\ket{-}$ with eigenvalues $\omega \pm gN^2$. Because transitions driven by baths $X$ (blue) and $Y$ (orange) become delocalized across all four eigenstates, multiple thermal short-circuit channels emerge (indicated by the circular arrows), bypassing the demon cycle and suppressing the reverse heat current.}
    \label{fig:nonbipar}
\end{figure}

The behavior changes qualitatively when the autonomous demon configuration is introduced, namely when subsystem $Y$ is coupled to two baths. In the bipartite model, this setup drives the intended four-state demon cycle and produces the  $O(N)$ reverse heat current discussed above. In the non-bipartite model, however, the same configuration fails. As illustrated in Fig.~\ref{fig:nonbipar}, because the exchange interaction does not commute with the Hamiltonians of the subsystems, the transition operators become delocalized across the hybridized eigenstates $\ket{\pm}$, opening multiple direct thermal short-circuit channels between the baths. These short-circuit channels bypass the demon cycle and suppress the reverse heat current, rendering the non-bipartite demon inoperable in the steady state unless additional energy shift or energy filtering is imposed to block such short-circuit pathways.

\subsection{Quantitative analysis of short-circuit currents in the non-bipartite model}

We now estimate the magnitude of the short-circuit currents in the non-bipartite demon configuration discussed in the previous subsection. We assume that bath $Y_2$ couples to the larger gap $\omega_+$, while bath $Y_1$ couples to the smaller gap $\omega_-$. We define the total upward and downward rates at each frequency as
\begin{align}
    U_+ &= \Gamma_{X\uparrow}^{\omega_+}+\Gamma_{Y_2\uparrow}^{\omega_+},
    &
    D_+ &= \Gamma_{X\downarrow}^{\omega_+}+\Gamma_{Y_2\downarrow}^{\omega_+},\\
    U_- &= \Gamma_{X\uparrow}^{\omega_-}+\Gamma_{Y_1\uparrow}^{\omega_-},
    &
    D_- &= \Gamma_{X\downarrow}^{\omega_-}+\Gamma_{Y_1\downarrow}^{\omega_-}.
\end{align}
The forward product of the four-state cycle
\(\ket g\to \ket+\to\ket e\to\ket-\to\ket g\) is
\begin{equation}
    U_+U_-D_+D_-,
\end{equation}
whereas the reverse product is
\begin{equation}
    U_-U_+D_-D_+.
\end{equation}
The affinity of this four-state cycle therefore vanishes:
\begin{equation}
    \mathcal A_{\rm cyc}
    =
    \ln
    \frac{U_+U_-D_+D_-}{U_-U_+D_-D_+}
    =
    0.
\end{equation}
After coarse-graining over the bath labels, this four-state network has only one independent cycle. Hence, the corresponding population cycle current vanishes:
\begin{equation}
    J_{\rm cyc}=0.
\end{equation}
The remaining currents are bath-resolved two-bath leakage currents. 
For the $\omega_+$ channel, the leakage current between $Y_2$ and $X$ is
\begin{equation}
    J_{\rm leak}^{(+)}
    =
    \frac{N^2}{2}
    \frac{
        \Gamma_{Y_2\uparrow}^{\omega_+}\Gamma_{X\downarrow}^{\omega_+}
        -
        \Gamma_{Y_2\downarrow}^{\omega_+}\Gamma_{X\uparrow}^{\omega_+}
    }{
        \Gamma_{X\uparrow}^{\omega_+}+\Gamma_{X\downarrow}^{\omega_+}
        +\Gamma_{Y_2\uparrow}^{\omega_+}+\Gamma_{Y_2\downarrow}^{\omega_+}
    } .
\end{equation}
We take $J_{\rm leak}^{(+)}>0$ to denote the current in which bath $Y_2$ excites the system and bath $X$ relaxes it. For the $\omega_-$ channel, the leakage current between $Y_1$ and $X$ is
\begin{equation}
    J_{\rm leak}^{(-)}
    =
    \frac{N^2}{2}
    \frac{
        \Gamma_{Y_1\uparrow}^{\omega_-}\Gamma_{X\downarrow}^{\omega_-}
        -
        \Gamma_{Y_1\downarrow}^{\omega_-}\Gamma_{X\uparrow}^{\omega_-}
    }{
        \Gamma_{X\uparrow}^{\omega_-}+\Gamma_{X\downarrow}^{\omega_-}
        +\Gamma_{Y_1\uparrow}^{\omega_-}+\Gamma_{Y_1\downarrow}^{\omega_-}
    } .
\end{equation}
We define $J_{\rm leak}^{(-)}>0$ analogously for the $\omega_-$ channel, with bath $Y_1$ exciting the system and bath $X$ relaxing it. The corresponding heat currents are
\begin{equation}
    \dot Q_{Y_2}^{\rm leak}
    =
    \omega_+ J_{\rm leak}^{(+)},
    \qquad
    \dot Q_{Y_1}^{\rm leak}
    =
    \omega_- J_{\rm leak}^{(-)}.
\end{equation}
Using local detailed balance,
\begin{equation}
    \frac{\Gamma_{X\downarrow}^{\omega_\pm}}{\Gamma_{X\uparrow}^{\omega_\pm}}
    =
    e^{\beta_X\omega_\pm},
    \qquad
    \frac{\Gamma_{Y_2\downarrow}^{\omega_+}}{\Gamma_{Y_2\uparrow}^{\omega_+}}
    =
    e^{\beta_{Y_2}\omega_+},
    \qquad
    \frac{\Gamma_{Y_1\downarrow}^{\omega_-}}{\Gamma_{Y_1\uparrow}^{\omega_-}}
    =
    e^{\beta_{Y_1}\omega_-},
\end{equation}
we obtain
\begin{align}
    J_{\rm leak}^{(+)}
    &=
    \frac{N^2}{2}
    \frac{
        \Gamma_{Y_2\uparrow}^{\omega_+}\Gamma_{X\uparrow}^{\omega_+}
        \left(
        e^{\beta_X\omega_+}
        -
        e^{\beta_{Y_2}\omega_+}
        \right)
    }{
        \Gamma_{X\uparrow}^{\omega_+}+\Gamma_{X\downarrow}^{\omega_+}
        +\Gamma_{Y_2\uparrow}^{\omega_+}+\Gamma_{Y_2\downarrow}^{\omega_+}
    },
    \\
    J_{\rm leak}^{(-)}
    &=
    \frac{N^2}{2}
    \frac{
        \Gamma_{Y_1\uparrow}^{\omega_-}\Gamma_{X\uparrow}^{\omega_-}
        \left(
        e^{\beta_X\omega_-}
        -
        e^{\beta_{Y_1}\omega_-}
        \right)
    }{
        \Gamma_{X\uparrow}^{\omega_-}+\Gamma_{X\downarrow}^{\omega_-}
        +\Gamma_{Y_1\uparrow}^{\omega_-}+\Gamma_{Y_1\downarrow}^{\omega_-}
    } .
\end{align}
Therefore, under the demon temperature ordering
\begin{equation}
    \beta_X>\beta_{Y_2}>\beta_{Y_1},
\end{equation}
both leakage currents are positive:
\begin{equation}
    J_{\rm leak}^{(+)}>0,
    \qquad
    J_{\rm leak}^{(-)}>0.
\end{equation}
Thus, heat is extracted from both $Y_2$ and $Y_1$ and dissipated into the colder bath $X$:
\begin{equation}
    \dot Q_{Y_2}^{\rm leak}>0,
    \qquad
    \dot Q_{Y_1}^{\rm leak}>0.
\end{equation}
This is opposite to the desired demon operation, where the hotter bath $Y_1$ should receive heat, i.e., $\dot Q_{Y_1}<0$.
In the linear-response regime, $\beta_X-\beta_{Y_1}=O(1/N), \quad\beta_X-\beta_{Y_2}=O(1/N)$, the leakage currents scale as
\begin{align}
    J_{\rm leak}^{(+)}
    &\simeq
    \frac{N^2}{2} G_+ \omega_+(\beta_X-\beta_{Y_2}),
    \\
    J_{\rm leak}^{(-)}
    &\simeq
    \frac{N^2}{2} G_- \omega_-(\beta_X-\beta_{Y_1}),
\end{align}
where $G_\pm=O(1)$ are positive finite conductance factors determined by the microscopic transition rates. 
The leakage current scales as 
\begin{equation}
    J_{\rm leak}^{(\pm)}
    =
    O(N^2\Delta\beta).
\end{equation}
In the near-equilibrium scaling used for the autonomous demon, $\Delta\beta=O(1/N)$, this becomes
\begin{equation}
    J_{\rm leak}^{(\pm)}=O(N),
    \qquad
    \dot Q_{Y_i}^{\rm leak}=O(N) \quad (i=1,2).
\end{equation}
Thus, the short-circuit currents are of the same order as the desired bipartite demon current. In the present symmetric non-bipartite model, however, the demon-cycle current vanishes, $J_{\rm cyc}=0$. The steady-state transport is therefore not a competition between a useful demon cycle and leakage cycles; only the leakage currents remain. Restoring demon operation would require breaking this symmetry, for instance by introducing an additional diagonal interaction such as $H_{\rm int}^{B}$ together with $H_{\rm int}^{NB}$. Such an interaction shifts the transition energies and can enable energy-selective filtering that suppresses the short-circuit channels.

\section{Bounds on the unitary information flow}

In this section, we derive Eq.~\eqref{unibound}, which is a general upper bound on the unitary contribution to the information flow. These bounds clarify how the possible scaling of the unitary information flow is controlled by the collective enhancement of the interaction Hamiltonian.

\subsection{Quantum-Fisher-information bound}

Let $\rho=\sum_n p_n \ket{n}\bra{n}$ be the spectral decomposition of the density matrix. For arbitrary Hermitian operators $X$ and $Y$, the following inequality holds:
\begin{equation}
    \left|
    \Tr\left(\rho[X,Y]\right)
    \right|^2
    \le
    \mathcal F_\rho(X)\,
    V_\rho(Y),
    \label{eq:qfi_commutator_bound}
\end{equation}
where
\begin{equation}
    \mathcal F_\rho(X)
    \coloneqq
    2\sum_{m\neq n}
    \frac{(p_n-p_m)^2}{p_n+p_m}
    |\langle n|X|m\rangle|^2
    \label{eq:qfi_def}
\end{equation}
is the quantum Fisher information, and
\begin{equation}
    V_\rho(Y)
    \coloneqq
    \Tr[\rho Y^2]
    -
    \Tr[\rho Y]^2
\end{equation}
is the variance of $Y$ with respect to $\rho$.
The unitary information flow is
\begin{equation}
    \dot I_Y^{\rm uni}
    =
    \Tr\left[
    i[H_{\rm int},\rho]\,
    \ln\rho_Y
    \right].
\end{equation}
Applying Eq.~\eqref{eq:qfi_commutator_bound} with
\begin{equation}
    X=H_{\rm int},
    \qquad
    Y=\ln\rho_Y,
\end{equation}
we obtain
\begin{equation}
    |\dot I_Y^{\rm uni}|^2
    \le
    \mathcal F_\rho(H_{\rm int})\,
    V_\rho(\ln\rho_Y).
\end{equation}
The variance term reduces to the local variance with respect to $\rho_Y$:
\begin{align}
    V_\rho(\ln\rho_Y)
    =
    V_{\rho_Y}(\ln\rho_Y).
\end{align}
Therefore, we arrive at
\begin{equation}
    |\dot I_Y^{\rm uni}|^2
    \le
    \mathcal F_\rho(H_{\rm int})\,
    V_{\rho_Y}(\ln\rho_Y).
    \label{eq:unitary_information_qfi_bound}
\end{equation}

\subsection{Derivation of the commutator bound~\eqref{eq:qfi_commutator_bound}}

For completeness, we provide the derivation of Eq.~\eqref{eq:qfi_commutator_bound}. The proof follows the method used in Ref.~\cite{sekiguchi2024improvementspeedlimitsquantum,yamauchi}. Using the spectral decomposition $\rho=\sum_n p_n \ket{n}\bra{n}$, we obtain
\begin{align}
    \Tr(\rho[X,Y])
    &=
    \sum_{m,n}
    (p_n-p_m)
    \langle n|X|m\rangle
    \langle m|Y|n\rangle .
\end{align}
Since the terms with $m=n$ vanish, the sum can be restricted to $m\neq n$. By the Cauchy--Schwarz inequality,
\begin{align}
    \left|
    \Tr(\rho[X,Y])
    \right|^2
    &\le
    \left[
    2\sum_{m\neq n}
    \frac{(p_n-p_m)^2}{p_n+p_m}
    |\langle n|X|m\rangle|^2
    \right]
    \left[
    \frac{1}{2}
    \sum_{m\neq n}
    (p_n+p_m)
    |\langle m|Y|n\rangle|^2
    \right].
\end{align}
The first factor is $\mathcal F_\rho(X)$. The second factor satisfies
\begin{align}
    \frac{1}{2}
    \sum_{m\neq n}
    (p_n+p_m)
    |\langle m|Y|n\rangle|^2
    &=
    \Tr[\rho Y^2]
    -
    \sum_n p_n |\langle n|Y|n\rangle|^2
    \nonumber\\
    &\le
    \Tr[\rho Y^2]
    -
    \left(
    \sum_n p_n \langle n|Y|n\rangle
    \right)^2
    \nonumber\\
    &=
    V_\rho(Y),
\end{align}
where we used Jensen's inequality in the second line. This proves Eq.~\eqref{eq:qfi_commutator_bound}.

\section{Transient dynamics in the bipartite model}
\subsection{Analytical calculation}
We analyze the bipartite model with collective interaction in the effective four-level subspace
$\{\ket{g},\ket{\psi_1},\ket{\psi_2},\ket{e}\}$. Unlike the interaction Hamiltonian used in the analysis of the autonomous Maxwell's demon model, the collective interaction used here has an effective strength $\lambda N^2$, so that the transition energies become $\Delta E_\pm=\omega\pm2\lambda N^2>0$.
For the bipartite interaction, the unitary dynamics does not change the local energy populations of subsystem $Y$ and hence
\begin{equation}
    \dot J_Y^{\rm uni}=
    \Tr\!\left[-i[H_{\rm int}^{B},\rho]H_Y\right]=0.
\end{equation}
It only generates a conditional phase rotation of the local coherence of $Y$. A direct calculation gives
\begin{equation}
    \dot I_Y^{\rm uni}(t)
=
\frac{4\lambda N^2}{r(t)}
\mathrm{Im}
\left[
\bigl(P_{\psi_1 e}(t)+P_{g\psi_2}(t)\bigr)
\bigl(P_{\psi_2 g}(t)-P_{e\psi_1}(t)\bigr)
\right]
\ln\frac{1+r(t)}{1-r(t)} ,
\end{equation}
where
\begin{equation}
    P_{ab}(t)\coloneqq\bra{a}\rho(t)\ket{b},
    \qquad a\neq b,
\end{equation}
and $r(t)$ is the Bloch-vector length of the reduced state $\rho_Y(t)$. 
\begin{equation}
    r(t)
    :=
    \sqrt{
    \left(P_{eY}(t)-P_{gY}(t)\right)^2
    +
    4|P_{egY}(t)|^2
    }.
\end{equation}
The reduced state of subsystem $Y$ in the local basis $\{\ket{g,Y},\ket{e,Y}\}$ is
\begin{equation}
    \rho_Y(t)
    =
    \begin{pmatrix}
        P_{gY}(t) & P_{geY}(t) \\
        P_{egY}(t) & P_{eY}(t)
    \end{pmatrix}
    =
    \begin{pmatrix}
        P_g(t)+P_{\psi_1}(t)
        &
        P_{g\psi_2}(t)+P_{\psi_1 e}(t)
        \\
        P_{\psi_2 g}(t)+P_{e\psi_1}(t)
        &
        P_{\psi_2}(t)+P_e(t)
    \end{pmatrix}.
\end{equation}
This expression explicitly shows that the unitary information flow can scale as $O(N^2)$ when the coherence factor remains $O(1)$.

The charging process is characterized by the nonequilibrium free energy $F_Y(t) = \Tr[H_Y\rho_Y] - \beta_Y^{-1}S(\rho_Y)$. For a general two-level state, define $r(t)=\sqrt{(P_{gY}(t)-P_{eY}(t))^2+4|P_{egY}(t)|^2}$ and let $p_\pm(t)=[1\pm r(t)]/2$ be the eigenvalues of $\rho_Y(t)$. Then
\begin{align}
    F_Y(t) &= \omega P_{eY}(t) + \beta_Y^{-1}\left[p_+(t)\ln p_+(t)+p_-(t)\ln p_-(t)\right],
\end{align}
where $P_{gY}$ and $P_{eY}$ are the ground- and excited-state populations of subsystem $Y$, and $P_{egY}$ is the off-diagonal coherence in the local energy basis. 

\subsection{Numerical calculation}

The transient dynamics presented in Fig.~\ref{fig:bipartitetrans} were obtained by numerically integrating the master equation above in the four-dimensional collective subspace $\{\ket{g}, \ket{\psi_1}, \ket{\psi_2}, \ket{e}\}$, using \\
\texttt{scipy.integrate.solve\_ivp} with the RK45 method. The density matrix $\rho(t)$ was evolved on a grid of 100 points in $t \in [0, 10]$ for the time-evolution plots (Fig.~\ref{fig:bipartitetrans}), with $N=10$.
The parameters were set to $\omega = 2.0$, $\lambda = 0.005$, $\beta_X = \beta_Y = 0.5$, and $\gamma_0 = 10^{-5}$. The transition rates satisfy local detailed balance as in the steady-state calculations. The initial state for the bipartite case is given in Eq.~\eqref{bipini}.
For the scaling analysis shown in Fig.~\ref{fig:biptransscale}, we used the same numerical procedure but evaluated the dynamics on a finer grid of $1000$ points in $t\in[0,50]$. The calculation was repeated for integer values of $N$ from $6$ to $19$. In this scaling analysis, we set $\lambda=0.002$ to keep the transition energies $\Delta E_\pm$ positive over the entire range of $N$ considered.

\section{Transient dynamics in the non-bipartite model}
\subsection{Analytical calculation}
We consider the non-bipartite collective interaction in Eq.~\eqref{nonbipint}. In the effective basis $\{\ket{e},\ket{+},\ket{-},\ket{g}\}$, the interaction couples the single-excitation states with matrix element $\Omega=gN^2$. The characteristic timescale of the coherent dynamics is therefore $O(1/N^2)$.
For the initial state used in Fig.~\ref{fig:nonbipartitetrans}, the density matrix remains block diagonal in this effective four-level basis, with coherence only between $\ket{+}$ and $\ket{-}$. The unitary information flow is
\begin{equation}
\dot I_Y^{\rm uni}
=
-2\Omega\mathrm{Im}[P_{\psi_1\psi_2}]
\ln\frac{P_{eY}}{P_{gY}}.
\end{equation}
When $\mathrm{Im}[P_{\psi_1\psi_2}]=O(1)$ and the logarithmic factor is $O(1)$, this gives
\begin{equation}
     \dot I_Y^{\rm uni}=O(\Omega)=O(N^2).
\end{equation}
Since $[H_{\rm int}^{NB},H_Y]\neq0$, the unitary energy flow $\dot J_Y^{\rm uni}$ is also nonzero and evolves on the same $O(1/N^2)$ timescale.
The local state $\rho_Y(t)$ is obtained by tracing out subsystem $X$. Since the contribution of $H_Y$ in the single-excitation subspace is $\omega\ket{\psi_2}\bra{\psi_2}$, the unitary energy flow is
\begin{align}
    \dot J_Y^{\rm uni}
    &=
    \Tr\!\left[-i[H_{\rm int}^{NB},\rho]H_Y\right]
    \nonumber\\
    &=
    2\omega\Omega\,
    \mathrm{Im}[P_{\psi_1\psi_2}],
\end{align}
where
$P_{\psi_1\psi_2}\coloneqq\bra{\psi_1}\rho\ket{\psi_2}$.
Thus, when the coherence $P_{\psi_1\psi_2}$ remains $O(1)$, the unitary energy flow scales as
\begin{equation}
    \dot J_Y^{\rm uni}=O(N^2).
\end{equation}
In this model, $\rho_Y(t)$ remains diagonal in the local energy basis:
\begin{equation}
     \rho_Y(t)=P_{gY}(t)\ket{g,Y}\bra{g,Y}
    +P_{eY}(t)\ket{e,Y}\bra{e,Y}.
\end{equation}
Thus, the nonequilibrium free energy reduces to
\begin{equation}
    F_Y(t)
    =
    \omega P_{eY}(t)+\beta^{-1}(P_{gY}(t)\ln P_{gY}(t)+P_{eY}(t)\ln P_{eY}(t))
\end{equation}
The nonequilibrium free energy shown in Fig.~\ref{fig:nonbipartitetrans} was evaluated from this reduced state.

\subsection{Numerical calculation}

The transient dynamics presented in Fig.~\ref{fig:nonbipartitetrans} were obtained by analytically evolving the density matrix $\rho(t)$ in the four-dimensional collective basis $\{\ket{g}, \ket{+}, \ket{-}, \ket{e}\}$. This calculation exploits the structure of the non-bipartite model, in which the populations and the coherence $\rho_{+-}$ decouple. The density matrix was evaluated on a grid of 100 points in $t \in [0.01,15]$. The thermodynamic parameters were set to $\omega = 2.0$, $g = 0.005$, $N = 10$ (so that $gN^2 = 0.5$), $\beta_X = \beta_Y = 0.5$, and the bare transition rate $\gamma_0 = 10^{-5}$. The initial state is given in Eq.~\eqref{nonbipini}.

\section{Dissipative information flow and energy-basis coherence}
\label{sec:dissipative_information_flow}
We first clarify how coherence in the energy eigenbasis enters the partial entropy production. Following the coherence-decomposition technique used in the proof of Supplementary Theorem~1 of Ref.~\cite{TajimaFuno}, we show that energy-basis coherence gives an additional non-negative contribution to entropy production. We then relate this coherence cost to the dissipative information flow.

\subsection{Entropy-production cost of energy-basis coherence}

We define the dephasing map
\begin{equation}
    \mathcal E[\chi]
    :=
    \sum_E
    \Pi_E
    \chi
    \Pi_E,
\end{equation}
and denote the energy-dephased state by
\begin{equation}
    \rho_d:=\mathcal E[\rho].
\end{equation}
To isolate the contribution from subsystem $X$, we introduce the corresponding short-time CPTP map
\begin{equation}
    \Lambda^X_{\Delta t}
    :=
    e^{\mathcal D_X\Delta t}
    =
    1+\mathcal D_X\Delta t+O(\Delta t^2).
\end{equation}
For the secular GKSL dynamics considered here, this map is dephasing-covariant in the energy eigenbasis:
\begin{equation}
    \Lambda^X_{\Delta t}\circ\mathcal E
    =
    \mathcal E\circ\Lambda^X_{\Delta t}.
\end{equation}
The relative entropy of coherence is
\begin{equation}
    C_{\rm rel}(\rho)
    :=
    D(\rho||\rho_d)
    =
    S(\rho_d)-S(\rho),
\end{equation}
where
\begin{equation}
    D(\rho||\sigma)
    :=
    \Tr[\rho(\ln\rho-\ln\sigma)] .
\end{equation}
By the monotonicity of quantum relative entropy under the CPTP map $\Lambda^X_{\Delta t}$, we obtain
\begin{align}
    C_{\rm rel}(\rho)
    -
    C_{\rm rel}(\Lambda^X_{\Delta t}[\rho])
    &=
    D(\rho||\mathcal E[\rho])
    -
    D\!\left(
    \Lambda^X_{\Delta t}[\rho]
    \middle\|
    \Lambda^X_{\Delta t}[\mathcal E[\rho]]
    \right)
    \nonumber\\
    &\ge 0.
\end{align}
Taking the limit $\Delta t\to0$, this implies
\begin{equation}
    \left.\frac{d}{dt}C_{\rm rel}(\rho) \right|_{t=0}
    \le 0.
\end{equation}
We now relate this monotonicity to the partial entropy production. 
Using the definition of the information flow, the partial entropy production can be written as
\begin{equation}
    \dot\sigma_X(\rho)
    =
    -\Tr[\mathcal D_X[\rho]\ln\rho]
    -
    \sum_{i_X}\beta_{i_X}\dot Q_{i_X}.
    \label{eq:sigma_X_simplified}
\end{equation}
The heat currents are invariant under dephasing between distinct energy eigenspaces
\begin{equation}
    \dot Q_{i_X}(\rho)
    =
    \dot Q_{i_X}(\rho_d).
\end{equation}
Consequently,
\begin{align}\label{eq:sigma_coherence_cost}
    \dot\sigma_X(\rho)-\dot\sigma_X(\rho_d)
    &=
    -\Tr[\mathcal D_X[\rho]\ln\rho]
    +
    \Tr[\mathcal D_X[\rho_d]\ln\rho_d]
    \nonumber\\
    &=
    -
    \left.\frac{d}{dt}C_{\rm rel}(\rho)\right|_{t=0}
    \nonumber\\
    &\ge 0.
\end{align}
This inequality shows that energy-basis coherence gives an additional non-negative contribution to the partial entropy production.

\subsection{Relation to the dissipative information flow}

The dissipative information flow is defined as
\begin{equation}
    \dot I_X^{\rm diss}
    =
    \Tr[\mathcal D_X[\rho]\ln\rho]
    -
    \Tr[\mathcal D[\rho]\ln\rho_X].
\end{equation}
By inserting $\ln\rho_d$, we decompose it as
\begin{align}
    \dot I_X^{\rm diss}
    &=
    \underbrace{
    \Tr\!\left[
    \mathcal D_X[\rho]
    \left(
    \ln\rho-\ln\rho_d
    \right)
    \right]
    }_{\dot I_{X,{\rm coh}}^{\rm diss}}
    +
    \underbrace{
    \Tr\!\left[
    \mathcal D_X[\rho]
    \left(
    \ln\rho_d-\ln\rho_X
    \right)
    \right]-\Tr\!\left[
    \mathcal D_Y[\rho]
    \ln\rho_X
    \right]
    }_{\dot I_{X,{\rm rem}}^{\rm diss}} .
    \label{eq:diss_info_decomposition_general}
\end{align}
The first term is the coherence-induced contribution. 
From Eq.~\eqref{eq:sigma_coherence_cost}, it is non-positive:
\begin{equation}
    \dot I_{X,{\rm coh}}^{\rm diss}
    =
    \Tr\!\left[
    \mathcal D_X[\rho]
    \left(
    \ln\rho-\ln\rho_d
    \right)
    \right]
    \le 0.
\end{equation}
Thus, energy-basis coherence gives a non-positive contribution to the dissipative information flow, or equivalently, a positive contribution to the partial entropy production.
Here, we consider states for which the global energy dephasing does not change the reduced state of subsystem $X$, namely $(\rho_d)_X=\rho_X$. Under this condition, the following simplification holds
\begin{equation}
    \dot I_X^{\rm diss}(\rho)
    \leq
    \dot I_X^{\rm diss}(\rho_d).
\end{equation}

% If your supplement is very short you might need to uncomment the following line to avoid
% layout problems with the figures and tables.
%\newpage

% \bibliography{ref.bib}  % moved/omitted in merged file to avoid a duplicate bibliography command
\end{document}